\documentclass[fleqn,usenatbib]{mnras}

\usepackage{newtxtext,newtxmath}

\usepackage[T1]{fontenc}

\DeclareRobustCommand{\VAN}[3]{#2}
\let\VANthebibliography\thebibliography
\def\thebibliography{\DeclareRobustCommand{\VAN}[3]{##3}\VANthebibliography}

\usepackage{graphicx}	
\usepackage{amsmath}	
\usepackage{bm} 

\title[MW mergers with SBI]{A simulation-based inference of the Milky Way merger history}

\author[A. Sante et al.]{
Andrea Sante,$^{1}$\thanks{E-mail: A.Sante@2022.ljmu.ac.uk}
Andreea S. Font, $^{1}$
Daisuke Kawata,$^{3,4}$
T. Lucas Makinen,$^{5,6}$
Robert J. J. Grand,$^{1}$
\\
$^{1}$Astrophysics Research Institute, Liverpool John Moores University, 146 Brownlow Hill, Liverpool L3 5RF, UK\\
$^{2}$Department, Institution, Street Address, City Postal Code, Country\\
$^{3}$Mullard Space Science Laboratory, University College London, Holmbury St Mary, Dorking, Surrey RH5 6NT, UK\\
$^{4}$National Astronomical Observatory of Japan, 2-21-1 Osawa, Mitaka, Tokyo 181-8588, Japan\\
$^{5}$Department of Applied Mathematics and Theoretical Physics, University of Cambridge, Wilberforce Rd, Cambridge, CB3 0WA, United Kingdom \\
$^{6}$Imperial Centre for Inference and Cosmology (ICIC), Imperial College London, Prince Consort Road, London SW7 2AZ, United Kingdom\\
}

\date{Accepted XXX. Received YYY; in original form ZZZ}

\pubyear{\the\year{}}

\begin{document}
\label{firstpage}
\pagerange{\pageref{firstpage}--\pageref{lastpage}}
\maketitle

\begin{abstract}
Accreted stars in the Milky Way preserve information about the progenitor galaxies in which they formed in their chemical and kinematic properties. In this study, we use the chemo-dynamical signatures in the merger debris to approximate the posterior distribution of the properties of disrupted satellites at their time of infall. Adopting a simulation-based inference framework, we train an ensemble of normalizing flows using samples of merger debris from the Auriga suite of simulations of Milky Way-like galaxies. Applying this methodology to a local sample of accreted stars in the Milky Way, we infer the lookback times, stellar and halo masses, as well as halo mass merger ratios of several known accretion events in the Galaxy: Gaia Enceladus-Sausage, Helmi streams, Heracles, I'itoi, LMS-1/Wukong, Sagittarius, Sequoia and Thamnos.

Our predictions align with the accretion time and mass estimates from the literature, as well as the expected relation between the progenitor stellar masses and debris metallicities across redshifts. The total stellar mass accreted from these events is predicted to be $2.2^{+1.1}_{-0.6}\times10^{9}~\rm{M_{\sun}}$, with the Gaia Enceladus-Sausage and Sagittarius dwarf galaxies being the largest contributors. The predicted stellar mass accreted from fully disrupted progenitors in the stellar halo is $1.3^{+1.0}_{-0.5}\times10^{9}~\rm{M_{\sun}}$, which is consistent with previous mass measurements of this component. We provide a prediction for the evolution of the Milky Way halo mass until the accretion of Sagittarius ($z\approx1$): specifically, we find that the mass growth of the Galaxy from the time of its first merger ($z\approx5$) to $z\approx2$ significantly exceeds the total mass of the known progenitors accreted during that interval, suggesting the presence of unidentified substructures. Our estimate of the Galaxy halo mass after the Sagittarius merger, but prior to the accretion of the Magellanic Clouds, is $5.9^{+1.4}_{-1.1}\times10^{11}~\rm{M_{\sun}}$.

\end{abstract}




\section{Introduction}

Within the hierarchical structure formation scenario, the Milky Way (MW) assembled its mass by accreting smaller satellite galaxies over time \citep{White_1978}. During these cosmic collisions, stars are tidally disrupted from the progenitor systems in which they formed, creating the stellar halo of the Galaxy \citep{Searle_1978, Johnston_1996}. Hence, this extended and diffuse component serves as a repository of fossil records of the events that describe the assembly history of the Galaxy \citep{Helmi_2020, Deason_2024, Bonaca_APW_2025}. 

The most apparent remnant of these merging events is the tidal tail of the Sagittarius dwarf galaxy \citep{Ibata_1994}, which appears as a stream of spatially coherent stars wrapping around a vast region in the Galaxy. Stellar streams are usually associated with recent accretion events, whose debris have not yet dynamically mixed with the stars that formed \textit{in-situ}, i.e., within the proto-MW. Signatures of more ancient accretion events can still be recovered as phase-mixed accreted stars share similar orbital properties in the \textit{integrals of motion} space \citep{helmi_1999}, which is defined by quantities expected to remain constant over time in a static gravitational potential. Leveraging this principle, \citet{helmi_1999} discovered the first evidence of a phase-mixed accretion event, which appeared as a substructure in the velocity space of stars in the solar neighbourhood. 

Because of the high precision required to separate stellar streams and accreted substructures from the dense background of the MW field stars, it was not until the second data release of the European Space Agency Gaia mission \citep{GaiaDR2} that it became possible to constrain the properties of the progenitors of the Galaxy. By providing accurate astrometry and proper motions for about a billion stars, fossil remnants of a plethora of merging events were uncovered from the stars in the inner halo \citep[for reviews see][]{Deason_2024, Bonaca_APW_2025}. The most significant detection was the identification of a large population of stars in highly radial and eccentric orbits, which represents the remnant of the Gaia-Enceladus-Sausage (GES) galaxy \citep{belokurov_2018, helmi_2018}, a massive dwarf galaxy believed to have been accreted around $8-10$ billion years ago. 

While the debris of the GES accretion event are expected to dominate the accreted stars population in the stellar halo \citep{Naidu_2020}, several smaller phase-mixed substructures have also been identified by complementing the Gaia catalog with radial velocity and chemical abundance measurements from high-resolution spectroscopic surveys, such as the Apache Point Observatory Galactic Evolution Experiment (APOGEE, \citealt{Majewski_2017}), the Galactic Archaeology with HERMES (GALAH, \citealt{Galah_2021}), the Hectochelle in the Halo at High Resolution
Survey (H3, \citealt{H3}) and the Large sky Area Multi-Object fiber Spectroscopic Telescope (LAMOST, \citealt{LAMOST_2012}). The application of clustering algorithms in this chemo-dynamical space has led to a proliferation of identified substructures, which include the retrograde Sequoia progenitor \citep{myeong_2019}, the low-energy retrograde structure Thamnos \citep{koppelman_2019_b} and the prograde LMS-1/Wukong stream \citep{Yuan_2020}. A further division of the retrograde halo into multiple populations with different metallicity has been proposed by \citet{Naidu_2020}, which includes the presence of the Arjuna and I'itoi substructures, which are, respectively, more metal rich and metal poor than the Sequoia debris.

Eventually, the search for the progenitors of the MW has extended into the inner regions of the Galaxy, where the debris of the most ancient mergers are expected to orbit. By analyzing the properties of globular clusters and field stars, several groups have postulated the existence of one or more massive, very early mergers \citep{Kruijssen_2020, horta_2021, Malhan_2024}. These events would have occurred more than 10-12 billion years ago and contributed to the formation of the proto-Galaxy. However, these debris are extremely phase-mixed and chemically similar to the primordial in-situ population \citep{belokurov_2023} making their definitive identification an ongoing challenge.  

Similarly, as the census of substructures in the stellar halo grows, so does the complexity of their interpretation. For instance, \citet{Horta_2023} analysed the detailed chemical abundance patterns of stars in these substructures and suggested that some of the retrograde systems, including Arjuna, I'itoi, and some definitions of Sequoia, are chemically indistinguishable from GES. This raises the possibility that, rather than being separate accretion events, these may be different dynamical components or fragments of a single, massive, and complex merger. Moreover, recent works have shown that stellar debris from a single progenitor can occupy a vast region of the chemo-dynamical space \citep{Jean-Baptiste_2017,Mori_2024} and clustering algorithms might return spurious or heavily contaminated substructures if not optimised \citep{Thomas_2025, Sante_2025}.  

While efforts are made to refine the identification of the individual progenitors of the MW, it is crucial to develop techniques for the characterisation of the merger events in order to piece together the MW assembly. Traditionally, the lookback time and mass of the progenitors at infall are estimated by fitting the properties of the respective debris with analytical models. For instance, the infall time of a disrupted galaxy can be inferred from the distribution of the stellar ages of its debris \citep{gallart_2019, montalban_2021, Woody_2025} as the star formation history is affected by the merger via the tidal stripping of cold gas. The stellar mass of progenitors can instead be determined by fitting the density profile of their debris \citep{Mackereth_2020,han_2022,Lane_2023}, or comparing the observed abundance of different chemical elements with the outputs of chemical evolution models \citep{Vincenzo_2019, Hasselquist_2021}. Alternatively, the properties of the progenitor galaxies at accretion can be estimated by comparing the observed chemo-dynamical distribution of their debris with those expected from specific mergers in cosmological simulations. 

While these methods provide interpretable estimates of the properties of progenitors, they are often fitted for single substructures following analytical assumptions (i.e., the isochrone model library for CMD fitting, the shape of the distribution function of the debris for the density profile fitting, or the choice of the chemical evolution model) that can be decoupled from the cosmological context in which the assembly occurs. Current cosmological simulations of MW-like galaxies provide a more comprehensive framework for the characterisation of mergers, as they self-consistently incorporate a wide range of coupled physical processes on both interstellar and cosmological scales.

One of the first studies to use the information encoded within cosmological simulations to reconstruct the assembly history of the MW was conducted by \cite{Kruijssen_2020}, who applied an artificial neural network to effectively map globular cluster properties (age and metallicity) in the E-MOSAICS suite \citep{Kruijssen_2019} onto the orbital parameters of the dwarf galaxies from which they originated. This methodology, combined with results from clustering algorithms, was then applied to the globular cluster population in the MW to recover single point estimates for the time of infall and stellar mass of the Galaxy progenitors. By treating the reconstruction of the merger tree as a regression task, potential degeneracies in the merger parameters were not accounted for in these predictions, which only reflected the uncertainty due to the configuration of the neural network.

In contrast, physical degeneracies can be explicitly accounted for by casting predictions of merger properties as a probabilistic inference problem using the mathematical framework of simulation-based inference \citep[SBI,][]{Cranmer_2020, Ho_2024}. SBI adopts generative machine learning techniques to directly approximate the likelihood or posterior distribution of intractable or unknown models from simulations. The main idea is to perform a Bayesian analysis by obtaining an approximation of the posterior distribution as informed by some observed realisations of the simulated model. Samples can then be drawn from this distribution effectively determining the most likely set of model parameters associated with the observations. \

One of the first applications of SBI to the reconstruction of the properties of disrupted mergers is the GalactiKit methodology \citep{galactikit}. By using the information contained in the merger histories from the 39 cosmological magnetohydrodynamical zoom-in simulations of MW-like galaxies in the Auriga suite \citep{Grand_2017, Grand_2024}, a model was developed to infer the lookback time, stellar mass, halo mass, and halo mass merger ratio at infall of disrupted dwarf galaxies directly from the present-day chemo-dynamical distribution of their debris.  

The aim of this paper is to apply the GalactiKit methodology to the accreted stars in the MW in order to provide a data-driven, self-consistent reconstruction of the assembly history of the Galaxy as depicted by the current census of its disrupted progenitors. For this analysis, we follow the data construction and division of stars in their respective accreted progenitor of origin defined in \cite{Horta_2023}. More details on the observational data used in this analysis are provided in Section~\ref{sec:data}. While the Auriga simulations reproduce morphological, chemical and dynamical trends observed in the stellar haloes of MW-mass galaxies \citep{Monachesi_2019}, there are still numerical offsets between the observed chemo-dynamical properties of stars and the ones of the particles in the simulations. In order to account for these discrepancies, we introduce a number of changes to the GalactiKit methodology, which are described in detail in Section~\ref{sec:method}. Finally, the chronology and properties of the merger events characterising the formation history of the MW as inferred through GalactiKit are reported in Section~\ref{sec:results1}. Section~\ref{sec:conclusions} presents our conclusions.


\section{Data}
\label{sec:data}
The chemo-dynamical properties for the stars considered in this analysis are obtained from a combination of the SDSS-IV/APOGEE \citep[DR17,][]{Majewski_2017, apogeedr17} and Gaia \citep[EDR3,][]{Gaia_edr3} surveys. The celestial coordinates and proper motions provided by Gaia are complemented by the radial velocities and chemical abundance measurements from APOGEE. The distance to each star is taken from the \texttt{astroNN} catalogue produced by \citet{leung_2019}, as it accounts for parallax zero-point biases in Gaia-EDR3 using photometric distance estimates from an artificial neural network model trained on APOGEE spectra. The 6-D phase-space coordinates are converted to a Galactocentric cylindrical coordinate system adopting the following solar parameters: a distance to the Galactic centre of $\mathrm{R}_{0}=8.178~\mathrm{kpc}$ \citep{GravityCollab_2019}, a vertical height above the Galactic mid-plane of $\mathrm{Z}_{0}=0.02~\mathrm{kpc}$ \citep{bennet_2019} and a velocity vector $[\mathrm{U_{\odot}}, \mathrm{V}_{\odot}, \mathrm{W}_{\odot}]=[-11.1,248.0,8.5]~\mathrm{km\,s^{-1}}$, which  accounts for the motion of the Sun relative to the Local Standard of Rest and Sgr A* \citep{Schonrich_2010, Reid_2020}.
The total energy of the stars is calculated assuming the \citet{McMillan_2017} model of the gravitational potential of the MW.

To ensure that the properties of the stars considered in the analysis are reliably measured and the samples of accreted debris of the disrupted progenitors are chemically distinct from each other and consistent with the dwarf satellite population of the MW, we define the accreted substructures in the Galaxy following the procedure described in \citet{Horta_2023}. Firstly, the following selection criteria are applied to the combined APOGEE-Gaia catalogue to identify the stars with the most reliable chemical abundance measurements:

\begin{itemize}
    \item $3500<T_{\mathrm{eff}}<5500~\mathrm{K} \; \mathrm{and} \; \log g<3.6,$
    \item APOGEE spectral S/N > 70,
    \item APOGEE $\mathtt{STARFLAG}=0$,
    \item $d_{\odot,\mathrm{err}}/d_{\odot}<0.2$,
\end{itemize}
where $T_{\mathrm{eff}}$ and $\log g$ are stellar atmospheric parameters determined through the APOGEE Stellar Parameter and Chemical Abundances Pipeline \citep[ASPCAP][]{GarciaPerez_2016}; APOGEE $\mathtt{STARFLAG}=0$ ensures that stars with issues in the ASPCAP fit are discarded, while $d_{\odot,\mathrm{err}}$ and $d_{\odot}$ are the relative uncertainty and value of the \texttt{astroNN}-derived distances to the Sun. Stars in the APOGEE Value Added
Catalogue of globular cluster stars \citep{Schiavon_2024} and those associated with the Large and Small Magellanic Clouds samples from \citet{Hasselquist_2021} are also excluded from the analysis. The resulting parent sample comprises 198,503 stars. 

Then, the debris from the disrupted progenitors of the MW are identified following the chemo-dynamical selection cuts described in Table~1 of \cite{Horta_2023}. However, we exclude from the analysis the stars from the Aleph and Nyx substructures as they are chemically consistent with the low-$\alpha$ and high-$\alpha$ disc populations, respectively. We also do not include stars from the Arjuna substructure, given their potential overlap with the high-energy retrograde debris from the GES merger \citep{naidu_2021,Horta_2023, Woody_2025}. Moreover, we do not consider the Icarus and Pontus substructures because they are comprised of a statistically insignificant number of stars in the parent sample ($<5$). The accreted substructures (and their respective number of stars in their sample) considered in this analysis are: Gaia-Enceladus-Sausage (2,380), the Helmi streams (104), Heracles (263), I'itoi (24), LMS-1/Wukong (89), Sagittarius (116), Sequoia and Thamnos (234).  The Sequoia sample is defined in three different ways following the selection criteria adopted in previous works by \citet{koppelman_2019}, \citet{myeong_2019} and \citet{Naidu_2020}. We follow the notation of \cite{Horta_2023} and refer to the samples selected through these selection criteria as \textit{K19} (41), \textit{M19} (165) and \textit{N20} (77), respectively. 

\begin{figure}
    \centering
    \includegraphics[width=\linewidth]{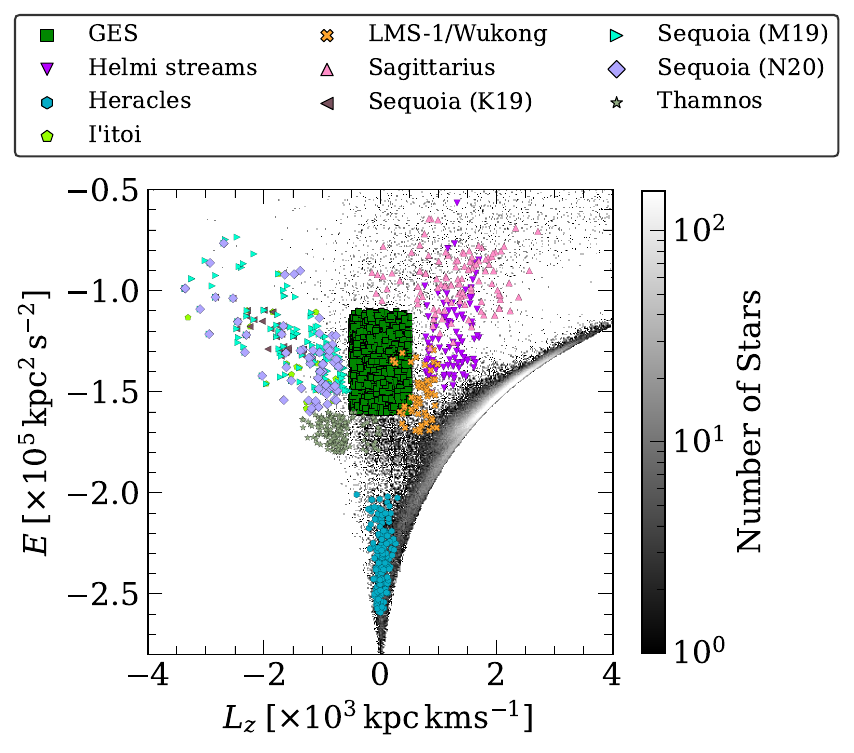}
    \caption{Distribution of the orbital properties ($E$ and $L_{z}$) of the accreted stars in the solar neighbourhood of the MW. The accreted stars are grouped by the proposed progenitor galaxy in which they formed, according to the criteria of \citet{Horta_2023}, and are plotted on top of the overall distribution of stars in the APOGEE sample from which they were selected.}
    \label{fig:IoM}
\end{figure}

\begin{figure*}
    \centering
    \includegraphics[width=0.8\textwidth]{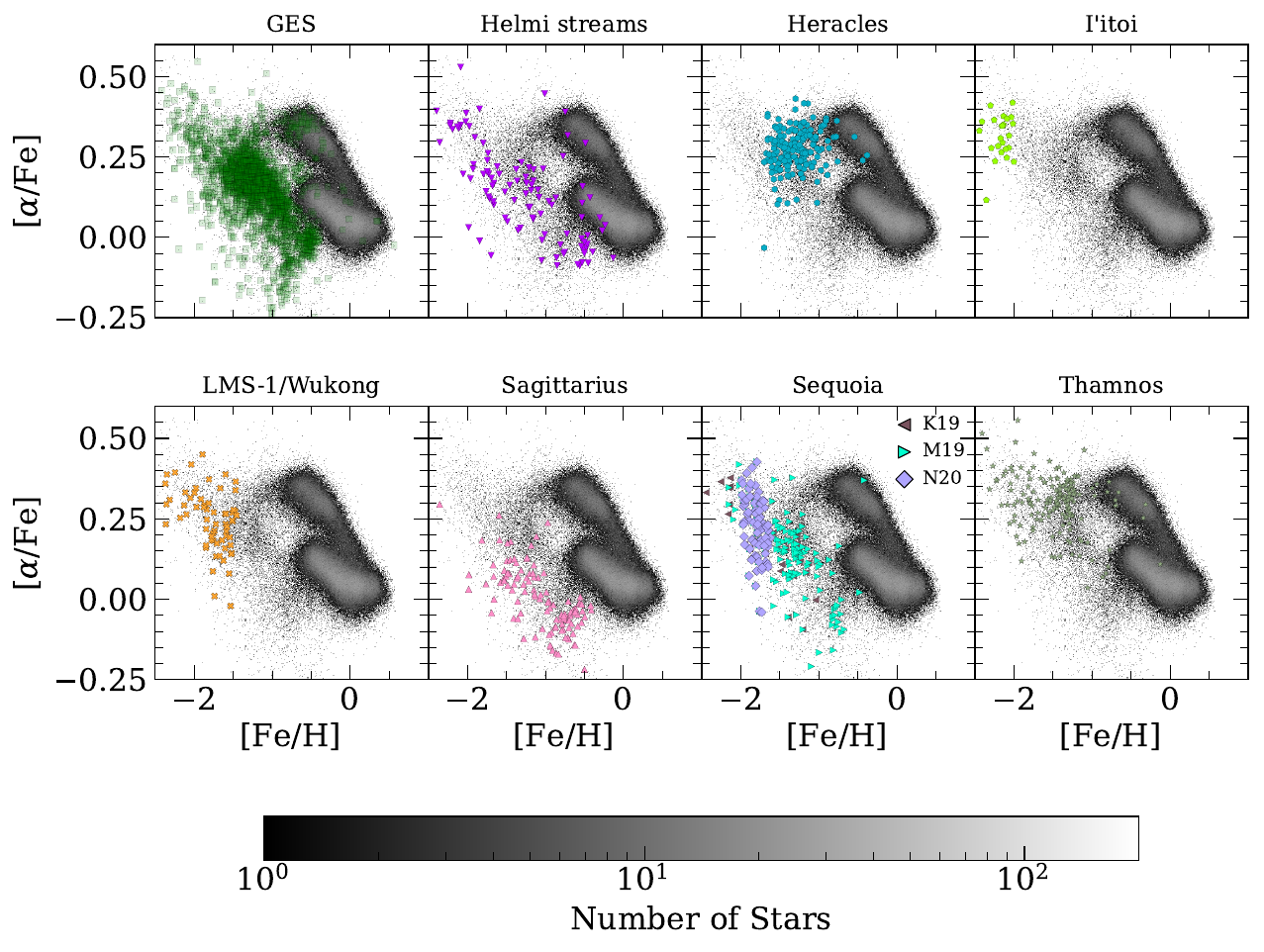}
    \caption{Chemical abundance ratios ([Mg/Fe] and [Fe/H]) distribution for the same samples of stars plotted in Fig.~\ref{fig:IoM}.}
    \label{fig:alphairon}
\end{figure*}

The chemo-dynamical properties of the stars in the accreted substructures considered in this study are reported in Figs.~\ref{fig:IoM} and \ref{fig:alphairon}, which show the distribution of debris in the integrals-of-motion and $\alpha$-iron planes, respectively. In each plot, stars belonging to the same substructures share the same colour and symbol.

\section{Inferring Merger Properties with Simulation-Based Inference}
\label{sec:method}
The primary objective of this work is to reconstruct the MW merger history by estimating the  properties of the past merger events. We approach this as a Bayesian inference problem, aiming to compute the posterior probability distribution for a set of progenitor parameters, $\bm{\theta}$, of an unknown model, $\mathcal{M}$, which determines the present-day chemo-dynamical properties, $\mathbf{x}$, of its stellar debris, i.e., $\mathcal{M}:\bm{\theta}\rightarrow\mathbf{x}$. 

An analytical definition of the posterior, $p(\bm{\theta}|\mathbf{x})\propto p(\bm{\theta})\,p(\mathbf{x}|\bm{\theta})$, is intractable as it would require an explicit formulation of likelihood probability, $p(\mathbf{x}|\bm{\theta})$, which accurately encapsulates the complex, coupled physical processes governing galaxy formation. These include the hierarchical assembly of galaxies within a cosmological framework, the distinct chemical evolution histories of the host and satellite, and the subsequent phase-mixing of accreted material within the host's evolving gravitational potential.

However, modern cosmological simulations, such as the Auriga suite \citep{Grand_2017, Grand_2024}, effectively forward-model these processes, providing a rich dataset of realistic merger events. As these simulations connect progenitor properties at the time of accretion to the corresponding stellar debris properties at redshift $z=0$, they are an implicit representation of the complex physical model, $\mathcal{M}$, that we want to approximate. 

The Auriga simulations are a suite of 39 magnetohydrodynamical (MHD) zoom-in simulations of MW-mass, relatively isolated, haloes selected from the L100N1504 \texttt{EAGLE} dark matter only simulation \citep{Schaye_2015}. The zoom-in simulations are run with the moving-mesh code \texttt{AREPO} \citep{Springel_2010} and include a comprehensive galaxy formation physics model, which is described in detail in \cite{Grand_2017}. Star particles are treated as single stellar populations following a Chabrier (\citeyear{Chabrier_2003}) initial mass function. Star formation is regulated by a pressurized, two-phase model for the interstellar medium (ISM) as gas cells become denser than 0.13 particles per cc. The chemical evolution is tracked by accounting for enrichment from Asymptotic Giant Branch (AGB) stars and both Type Ia and Type II supernovae. Feedback processes are also included as galactic-scale gaseous outflows powered by supernovae winds and active galactic nuclei feedback in both quasar and radio modes. Additionally, the model includes  primordial and metal-line gas cooling with self-shielding corrections, as well as a unidirectional, homogeneous magnetic field of $10^{-14}~\mathrm{G}$.

The combination of these physical prescriptions results in a population of simulated galaxies that successfully reproduce a wide range of observational scaling relations found in the stellar haloes of nearby MW analogues \citep{Monachesi_2019}. 

We extract the information in the Auriga simulation with Simulation-Based Inference \citep[SBI,][]{Cranmer_2020, Ho_2024}, a class of statistical methods that uses simulations to perform Bayesian inference when the likelihood is unknown or computationally prohibitive. In particular, we base our approach on GalactiKit, an SBI framework developed by \cite{Sante_2025} specifically to reconstruct the merger properties. The core idea of GalactiKit is to use an ensemble of Masked Autoregressive Flows \citep[MAFs,][]{Papamakarios_2017}, a type of generative artificial neural network capable of learning complex, high-dimensional probability distributions, to approximate the posterior distribution of the merger parameters directly from the joint distribution of the merger-debris pairs contained in the Auriga simulations.

Following the GalactiKit framework, we define the vector of merger parameters $\bm{\theta}$, composed of the quantities of interest to be inferred, i.e., the lookback infall time $\tau$, stellar mass $\log (M_*/\mathrm{M_{\sun}})$, halo mass $\log (M/\mathrm{M_{\sun}})$ and the halo mass merger ratio $\log (\mathrm{MMR})$ at infall. The data vector $\mathbf{x}$, which is used to inform the prediction of the posterior distribution, is defined as the  chemo-dynamical properties of 100 star particles randomly selected from the present-day ($z=0$) debris of the merger. For each star particle, we consider the total energy ($\log (E)$), total specific angular momentum ($\log (L)$), iron-to-hydrogen abundance ($[\mathrm{Fe}/\mathrm{H}]$), and $\alpha$-to-iron abundance ($[\alpha/\mathrm{Fe}]$).  From each merger event in the Auriga simulations, we draw multiple 100-star samples to augment the dataset, with a maximum of 10 samples per event to ensure that individual mergers with vast amounts of debris do not dominate the training process. In total, this sampling strategy yields 20,433 unique merger-debris pairs, 80\% of which are randomly selected for training the posterior estimator and the rest used for validating its performance.

The original GalactiKit framework was designed for application to simulated data. Therefore, the methodology needs to be adapted to account for potential discrepancies between the simulations and observations before being applied to real observations of the MW. These discrepancies can arise from a wide range of sources, such as the systematic differences between the subgrid physics models in Auriga and the true physical processes of the MW (e.g., chemical yield tables, feedback models) or the uncertainties inherent in the measurement of stellar properties.

\subsection{A calibration model for generating realistic training data}
\label{sec:noise_model}

Because the real deviations between the stellar properties in the simulations and observations, $\mathbf{\Delta x}$, are unknown, we assume that they can be described by an underlying stochastic model, $\mathcal{C}: \bm{\theta_{\mathcal{C}}}\to \bm{\Delta x}$. 
 
By perturbing the same debris sample with offset values drawn from $\mathcal{C}$, i.e., $\mathbf{\hat{x}=x+\Delta x}$, a new training dataset of merger-debris pairs, $(\bm{\theta}, \bm{\hat{x}})$, can be created. As each debris configuration, $\bm{x}$, is represented by multiple realisations with different calibration values, $\bm{\hat{x}}$, a posterior density estimator trained on this new dataset learns to estimate the posterior distribution, $p(\bm{\theta}|\bm{\hat{x}})$, of the merger parameters marginalising over the differences between simulations and observations. 

\begin{figure}
    \centering
    \includegraphics[width=0.5\linewidth]{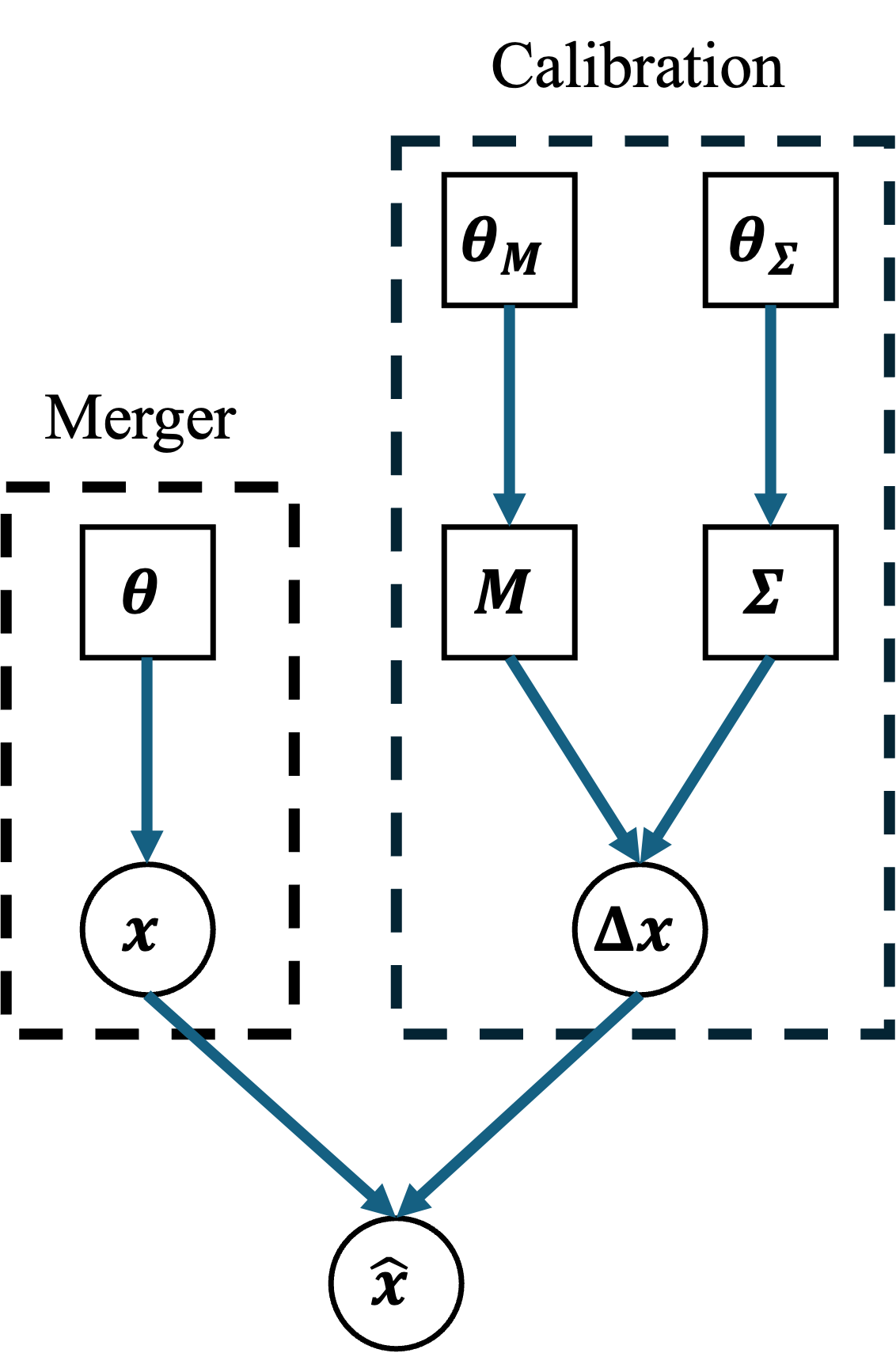}
    \caption{Relations between data, $\bm{x}$,  and parameters, $\bm{\theta}$ for the the model, $\mathcal{M}$, describing the relation between merger and debris properties, and the calibration model, $\mathcal{C}$, describing the difference between simulated and observed stellar properties. The parameters referring to each model are indicated by dashed boxes. }
    \label{fig:bayesian_model}
\end{figure}

A schematic of the relations between the merger-debris model, $\mathcal{M}$, and the calibration model, $\mathcal{C}$, is shown in Fig~\ref{fig:bayesian_model}. The actual $z=0$ chemo-dynamical distribution of debris $\mathbf{x}$ is parametrised by the properties of the associated disrupted progenitor galaxy, $\bm{\theta}$, while the observed distribution, $\bm{\hat{x}}$, is perturbed by the stochastic noise component, $\mathbf{\Delta x}$. This is modeled as a multivariate Gaussian distribution with mean and standard deviation vectors, $\mathbf{M}\,\text{and}\, \mathbf{\Sigma}$. For each $\mathbf{\Delta x}$, the values of the $i$-th components in $\mathbf{M}$ are drawn by a normal distribution, $\mathcal{N}_{i}(\mu_{\rm{M_i}},\sigma^{2}_{\rm{M}_i} )$.
The parameters $(\mu_{\rm{M_i}},\sigma^{2}_{\rm{M}_i})$ are chosen to reflect the degree of discrepancy between the simulated and observed debris properties. For instance, the gravitational potential and Solar motion are assumed to be reasonably well-modeled. Thus, the parameters of the calibration distribution for these properties are drawn from a Gaussian centred at zero, with a standard deviation of 0.10 to account for minor model inaccuracies and observational error, i.e., $\mu_{\log(E)}=\mu_{\log(L)}=0$ and $\sigma^2_{\log( E)}=\sigma^{2}_{\log(L)}=0.10$.

A more significant difference between the simulations and observations is noticed for the chemical abundance ratio distributions. Comparing the overall $[\mathrm{Fe}/\mathrm{H}]$ and $[\alpha/\mathrm{Fe}]$ distributions for the star particles in each of the 39 Auriga simulations to the ones for the stars in the MW, we find the former are systematically less $\alpha$-enhanced and more iron-rich than the latter. This could be related to several factors related to the subgrid physics implemented in the Auriga simulations, such as inaccuracies in the chemical yields tables or secondary processes unaccounted by the stellar evolution or feedback models used in the simulations. To approximately quantify the effect of these discrepancies, we calculate the difference between the peak in the chemical abundance ratio distributions of each Auriga simulation and the MW, $[\mathrm{Fe}/\mathrm{H}] _{\rm{MW}} -   [\mathrm{Fe}/\mathrm{H}]_{\rm{Auriga},\it{i}}$, where $i$ indicates one of the 39 simulations considered in the analysis. The mean and variance values for the calibration model are then derived from the median and variance of the set of these differences, i.e., $\mu_{[\mathrm{Fe}/\mathrm{H}]}=\rm{med}( [\mathrm{Fe}/\mathrm{H}] _{\rm{MW}} -   [\mathrm{Fe}/\mathrm{H}]_{\rm{Auriga},\it i})  =-0.20$, $\mu_{[\rm{Mg}/\mathrm{Fe}]}=\rm{med}( [\rm{Mg}/\mathrm{Fe}]_{\rm{MW}} -   [\rm{Mg}/\mathrm{Fe}]_{\rm{Auriga}, \it i} )=0.42$, $\sigma^2_{[\mathrm{Fe}/\mathrm{H}]}=\rm{var}( [\mathrm{Fe}/\mathrm{H}] _{\rm{MW}} -   [\mathrm{Fe}/\mathrm{H}]_{\rm{Auriga},\it i})=0.08^2$ and $\sigma^2_{[\rm{Mg}/\mathrm{Fe}]}=\rm{var}( [\rm{Mg}/\mathrm{Fe}]_{\rm{MW}} -   [\rm{Mg}/\mathrm{Fe}]_{\rm{Auriga}, \it i} )=0.02^2$.

To ensure the calibration model is realistic, we assume there is an unknown underlying correlation across the dimensions of the calibration model, hence avoiding to drawn the offset values from each component independently. This is achieved by sampling the correlation matrix of the model, $\mathbf{C}$, from a Lewandowski-Kurowicka-Joe (LKJ, \citeyear{LKJ_prior}) distribution  with a concentration parameter $\eta=1$. The covariance matrix of the calibration model is then calculated as $\mathbf{\Sigma} = \mathbf{S} \cdot \mathbf{C} \cdot \mathbf{S}$, where $\mathbf{S}$ is a diagonal matrix that contains the standard deviation of each stellar property. The values of the non-zero elements in $\mathbf{S}$ are drawn from an exponential distribution, which describes the internal spread in the noise assigned to each star in the debris sample associated with a specific merger event. This is assumed to be marginal compared to the values of the mean vector, $\mathbf{M}$. 

\begin{figure}
    \centering
    \includegraphics[width=\linewidth]{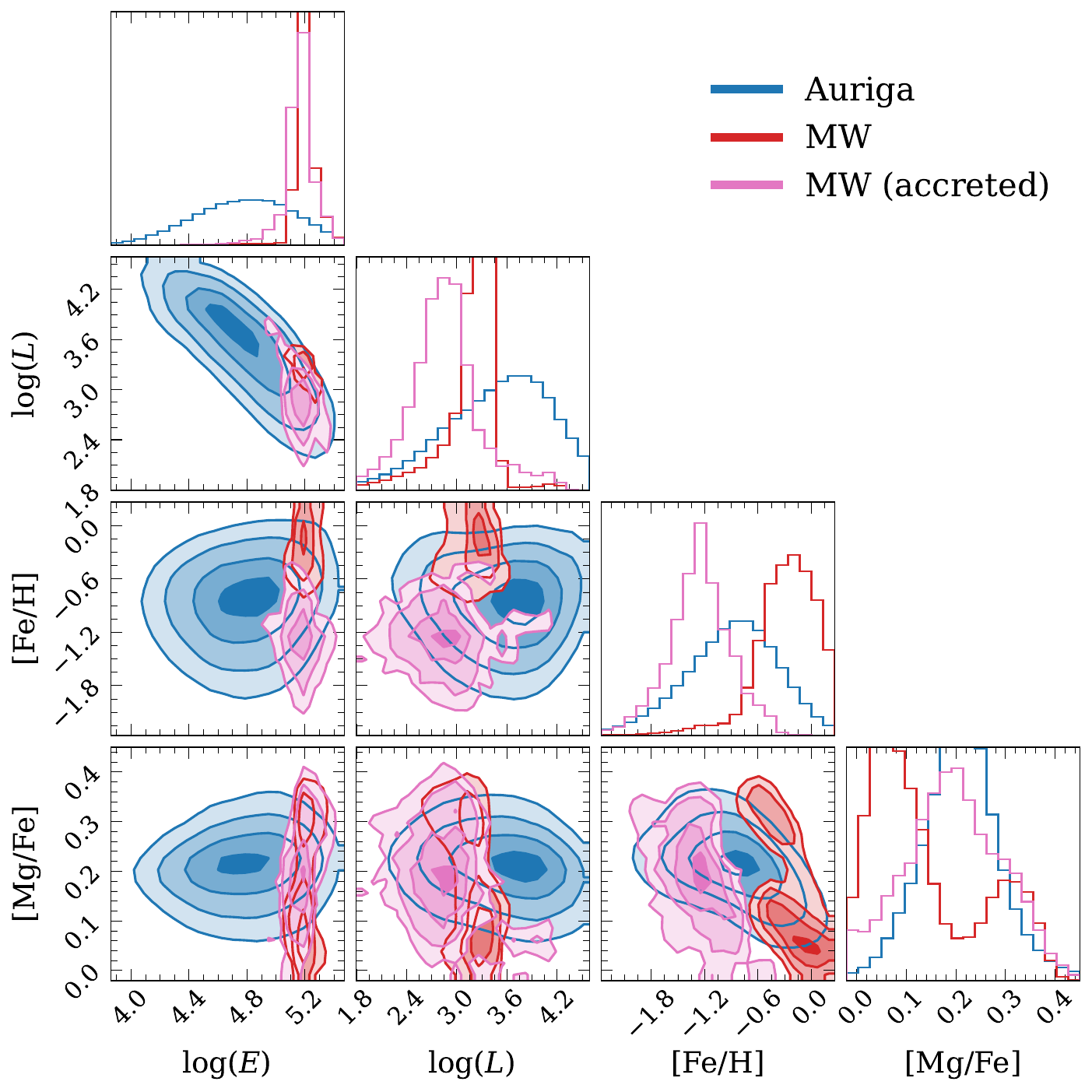}
    \caption{Chemo-dynamical properties of the accreted star particles from the Auriga simulations used to train the posterior estimator (blue) and of the stars in the MW parent sample obtained from the APOGEE and Gaia surveys (red). The distribution of the MW stars belonging to the accreted substructures considered is also shown (pink). The distributions are normalised by the total amount of samples.}
    \label{fig:datasets}
\end{figure}

A comparison between the chemo-dynamical properties of the accreted star particles in the training dataset as taken from the Auriga simulations and perturbed by the stochastically generated noise (blue) and the ones of the stars in the MW parent sample (red) is shown in Fig.~\ref{fig:datasets}. The stellar distributions are plotted as contours showing the 0.118, 0.393, 0.675 and 0.864 confidence levels. A selection of accreted stars (pink) is derived from the MW parent sample by considering all the stars in the accreted substructure samples.

\subsection{Data aggregation with Fishnets}
\label{sec:fishnets}

Along with the introduction of a stochastic noise model, a data aggregation step is also added to the original method developed in \citet{galactikit} before the data are passed to the MAFs ensemble for the posterior estimation. By compressing the debris data, $\bm{\hat{x}}$, into a lower dimensional representation, $\bm{\hat{x}_{\mathrm{agg}}}$, the inference becomes more robust to the specific values of the properties of the individual stars in the sample. For this purpose, we employ the Fishnets model \citep{Makinen_2023}, which is designed to produce robust, information-optimal embeddings from heterogeneous, noisy sets of data. The Fishnets framework is based on the additive nature of the Fisher information matrix and the log-likelihood score for independent data samples. The model operates via a pair of twin neural networks that learn to approximate the per-star score vector ($\mathbf{t}_i = \nabla_\theta \ln p(\bm{x}_i|\bm{\theta})$) and the Fisher information matrix ($\mathbf{F}_i = -\mathbb{E}_\theta[\nabla\nabla^T \ln p(\bm{x}_i|\bm{\theta})]$) for each star $i$ in a debris set.

For a set of $N$ stars comprising a single merger's debris, the total score and Fisher matrix are obtained by simple summation:
\begin{equation}
    \mathbf{t}_{\text{total}} = \sum_{i=1}^{N} \mathbf{t}_i \quad ; \quad \mathbf{F}_{\text{total}} = \sum_{i=1}^{N} \mathbf{F}_i
\end{equation}
These aggregated quantities, which can be combined to form a Maximum Likelihood Estimate of the parameters  ($\bm{\hat{x}}_{\mathrm{agg}} = \bm{\hat{\theta}} \propto \mathbf{F}_{\text{total}}^{-1}\mathbf{t}_{\text{total}}$), serve as an information-optimal summary statistic of the entire set of stellar debris. This summary is then used as an input to the posterior estimator.

The Fisher matrix acts as an optimal weighting scheme for each star. This makes the aggregation resilient to changes in the underlying data distribution, such as those introduced by the stochastic calibration model, ensuring that the inference is not biased by these variations and that the constraining power of the data is maximally preserved.

\subsection{Posterior density estimation and validation}
\begin{figure}
    \centering
    \includegraphics[width=0.65\linewidth]{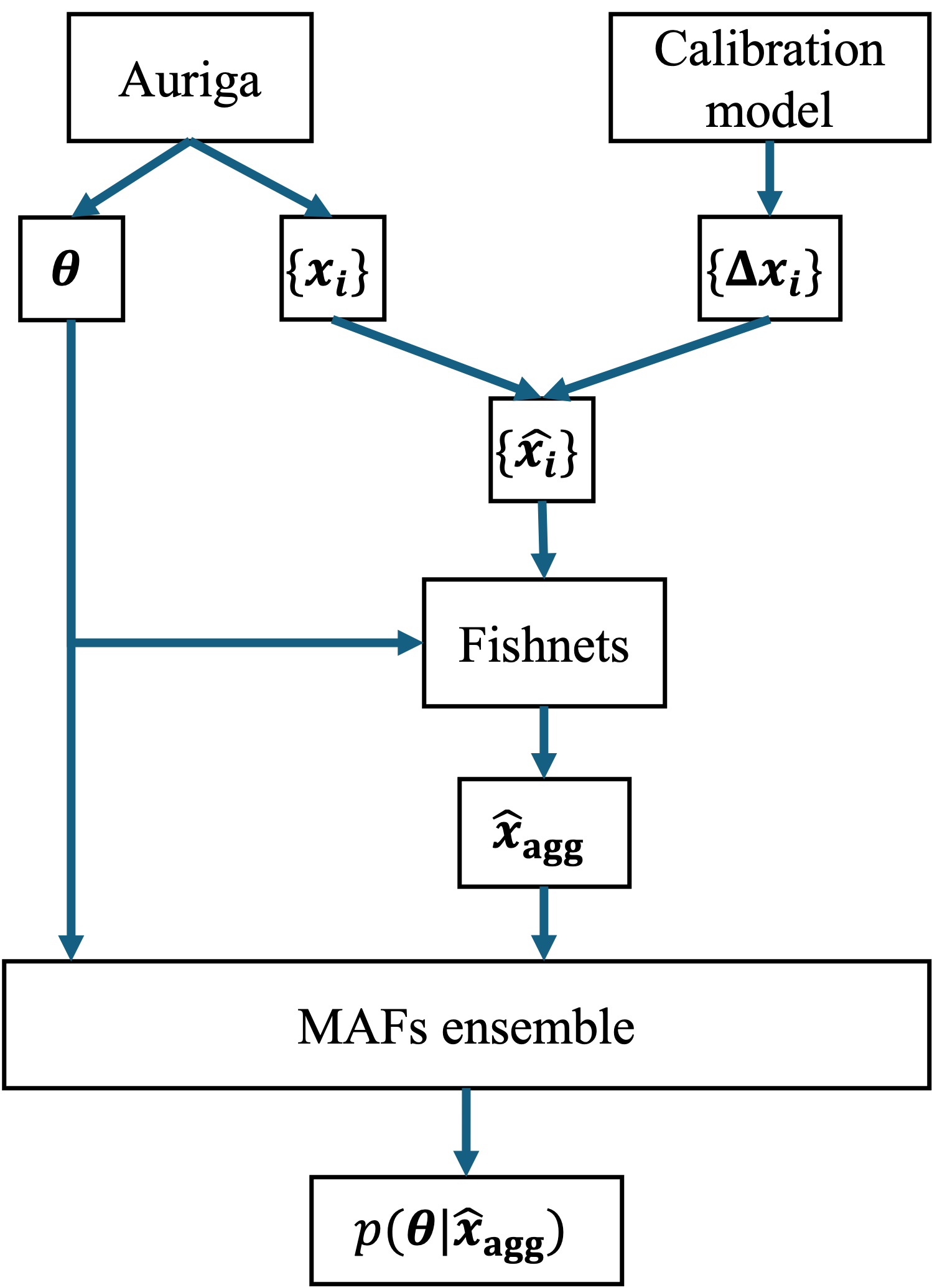}
    \caption{Outline of the pipeline used for the modelling of the posterior distribution of the properties of merger events, $\bm{\theta}$, from the observed $z=0$ chemo-dynamical properties of their debris, $\bm{\hat{x}}$.}
    \label{fig:pipeline}
\end{figure}

An end-to-end overview of the pipeline adopted to approximate the posterior probability of the properties of the disrupted progenitors of the MW is shown in Fig.~\ref{fig:pipeline}. At first, merger-debris pairs, representing the ground-truth merger parameters, $\bm{\theta}=(\tau/\rm{Gyr}, \log (M_*/\mathrm{M_{\sun}}), \log (M/\mathrm{M_{\sun}}), \rm{MMR} )$, and the chemo-dynamical data of the resulting stellar debris, $\mathbf{x}=(\log(E), \log(L),\rm{[Fe/H], [Mg/Fe]})$, are extracted from the Auriga simulations. To simulate realistic observational conditions, we generate 10,000 unique realisations, $\mathbf{\Delta x}$, from the calibration model. Data vectors are generated by randomly adding  calibration realisations to the vectors of stellar debris.

A compressed summary representation $\bm{\hat{x}_{\mathrm{agg}}}$ is obtained using a Fishnets model with 4 layers of 128 neurons each. To ensure robustness, a different, randomly selected calibration realization is applied to the debris data for each training epoch. This forces the model to learn the underlying features of the merger debris invariant to the specific calibration instance. The debris data is standardized using the median and inter-quartile range of the combined MW stellar sample. The Fishnets model is trained for 10,000 epochs with a batch size of 256 and a learning rate of 0.0001 using the Adam optimization algorithm \citep{kingma_2017}.

Once the Fishnets model is trained, we use it to generate the dataset of compressed data, $\bm{\hat{x}_{\mathrm{agg}}}$, which is used for the posterior density estimation. This dataset comprises ten examples of each merger-debris pair in the Fishnet training set, where the stellar properties are perturbed each time by a different calibration realisation. The posterior distribution of the merger parameters given the aggregated chemo-dynamical information of the debris, $p(\bm{\theta}|\bm{\hat{x}_{\mathrm{agg}}})$, is approximated by an ensemble of three MAFs.  Given the reduced dimensionality of the input data after the Fishnets compression, we simplified the MAF architecture to 10 layers of 100 neurons each. Each MAF is trained for 5,000 epochs, using the same optimizer, batch size, and learning rate as the Fishnets model. For each merger property, the associated posterior is derived assuming a uniform prior distribution spanning all the values expected for a merging dwarf galaxy, i.e., $p(\tau/\rm{Gyr})=\mathcal{U}(0,14)$, $p(\log (M_*/\mathrm{M_{\sun}}))=\mathcal{U}(6,11)$,  $p(\log (M/\mathrm{M_{\sun}}))=\mathcal{U}(8,12)$ and $p(\rm{MMR})=\mathcal{U}(-3,0)$.
The implementation and training of the density estimator models is done using the Learning the Universe
Implicit Likelihood Inference (LtU-ILI) framework \citep{Ho_2024}
with the \texttt{sbi} backend.

The goodness of the inference of the trained model is estimated by comparing the predicted posterior distributions to the true merger parameters for all examples in the validation set. Each validation example is perturbed by a calibration realisation not used in the training of the models, and then passed to the Fishnets model to obtain an aggregated representation. This is in turn used to estimate the posterior distribution of the merger properties with the MAFs ensemble. We sample the predicted posterior distribution 1,000 times per validation example to compute the median and the 16th-84th percentile range.

\begin{figure*}
    \centering
    \includegraphics[width=0.8\linewidth]{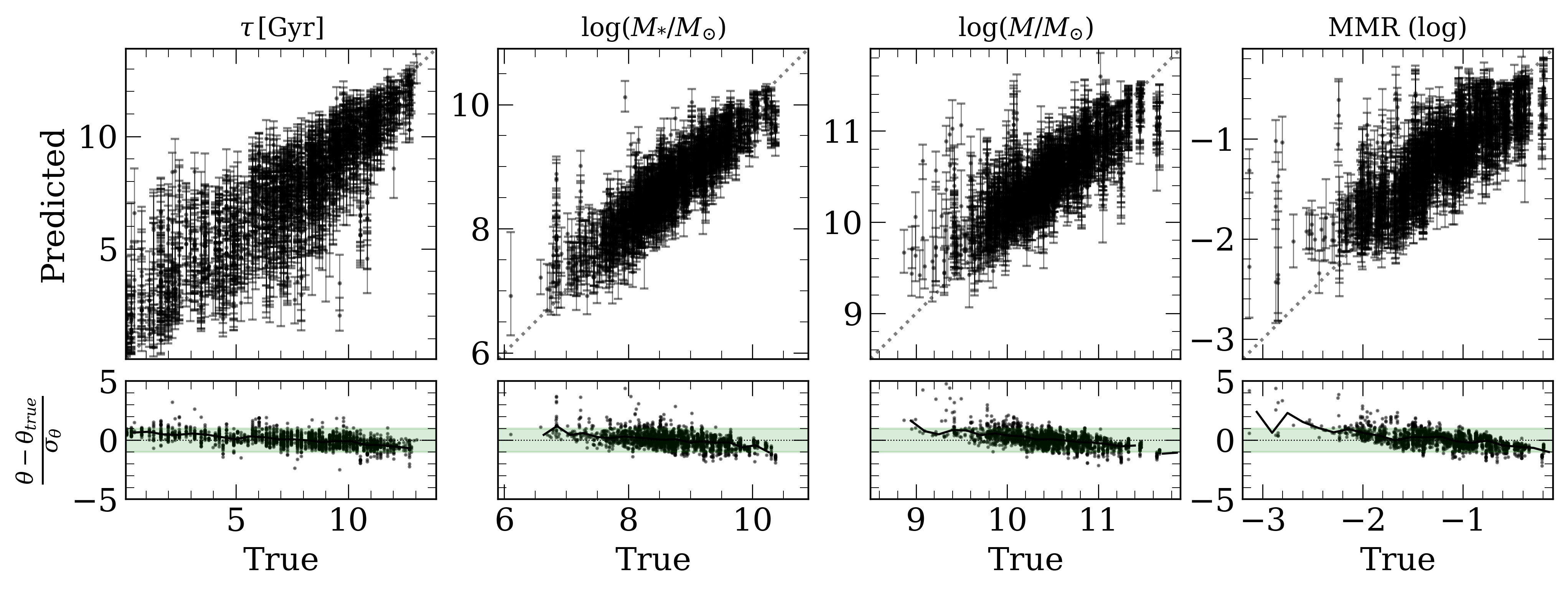}
    \caption{Predicted vs. true parameters for the validation set. The top panels show the median of the predicted posterior ($y$-axis) against the true value ($x$-axis). Error bars represent the 16th-84th percentile range. The bottom panels show the normalized residuals, with the green shaded region indicating the $\pm1$ deviation range.}
\label{fig:validation_scatter}
\end{figure*}

Fig.~\ref{fig:validation_scatter} shows a direct comparison between the true merger property values and the median of our predictions for the unseen Auriga mergers. The error bars indicate the predicted 16th-84th percentile range. The bottom panels display the normalized residuals, showing the difference between the predicted median and the true value, scaled by the 16th-84th percentile range. Overall, the predictions align well with the true values across the entire parameter range, with the residuals scattered around zero, indicating that there is no significant systematic bias.

\begin{figure*}
    \centering
    \includegraphics[width=0.8\linewidth]{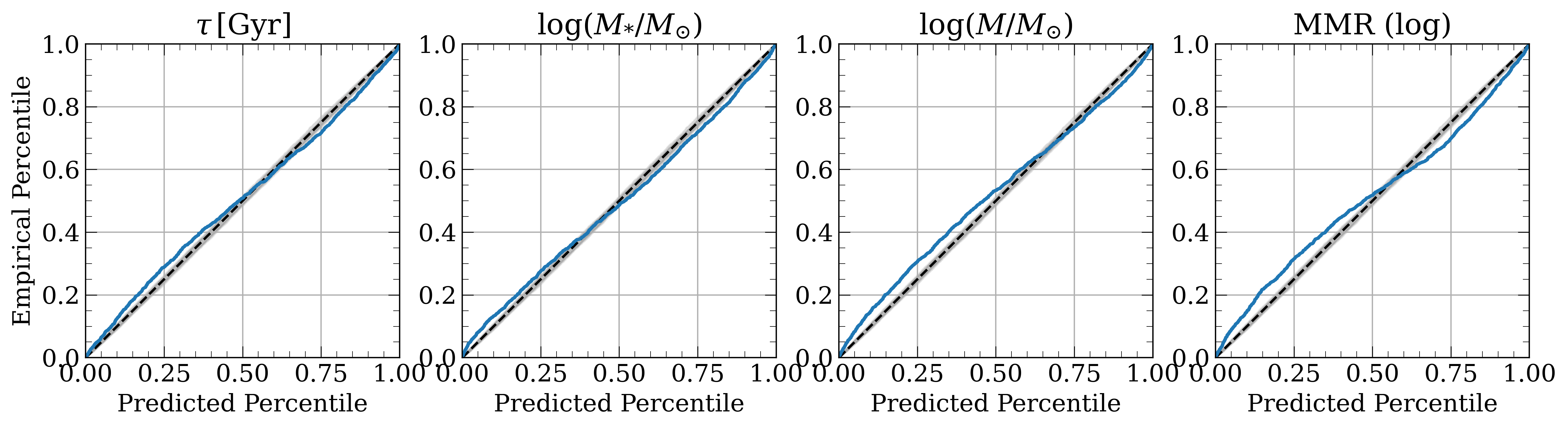}
    \caption{Validation of posterior calibration using the TARP method. The plot shows the empirical coverage probability (y-axis) versus the predicted credibility level (x-axis). The dashed black line represents a perfect estimator.}
    \label{fig:validation_tarp}
\end{figure*}

Moreover, the statistical accuracy of the posterior estimation is assessed with the Tests of Accuracy with Random Points \citep[TARP,][]{Lemos_2023} method. This test compares the expected coverage probability to the actual credibility levels of the estimated distribution. For an accurate posterior estimation, the $n\%$ credibility region should contain the true merger parameters $n\%$ of the time. The results of the TARP test, shown in Fig.~\ref{fig:validation_tarp}, reveal that the model is generally well-calibrated, but with some parameter-specific behaviors. While the calibration for infall time and stellar mass is good, with the empirical coverage probability following closely the ideal one-to-one relationship, the one for the total progenitor mass and the merger-mass ratio show more significant deviations. In both cases, the curves lie slightly above the diagonal for low credibility levels and dip slightly below for high ones. This indicates that the model is mildly under-confident in its narrower credible intervals (e.g., the 20\% credible region contains the true value more than 20\% of the time), meaning that the predicted uncertainties for the parameters are, on average, larger than they need to be. While this is preferable to overconfidence, it suggests that the model could be more precise in its predictions for this specific property.

\section{The time of accretion and stellar mass of the MW disrupted progenitors}
\label{sec:results1}

The chemo-dynamical properties of the accreted stars from the MW progenitors are used as conditions to the posterior estimation model to infer the lookback infall time, stellar mass, halo mass and halo mass merger ratio of accreted galaxies at infall. Given that the model can be conditioned only by the properties of a hundred stars, we implement a resampling strategy to ensure a robust and unbiased estimate of the posterior distribution to the selection of the debris sample. For each accreted substructure, we randomly select $n=\left \lceil{N/100}\right \rceil \times10 $ samples of 100 stars each, where $N$ is the total number of accreted stars in that specific progenitor sample. Each of these $n$ 
star distributions is then used to obtain an approximation of the posterior distribution of the merger properties. The resulting $n$ independently generated posteriors are sampled 1,000 times. For each progenitor, the summary statistics describing the properties of the galaxy at infall are calculated from the aggregated $n\times1,000$ posterior samples. 

\begin{figure*}
    \centering
    \includegraphics[width=0.6\linewidth]{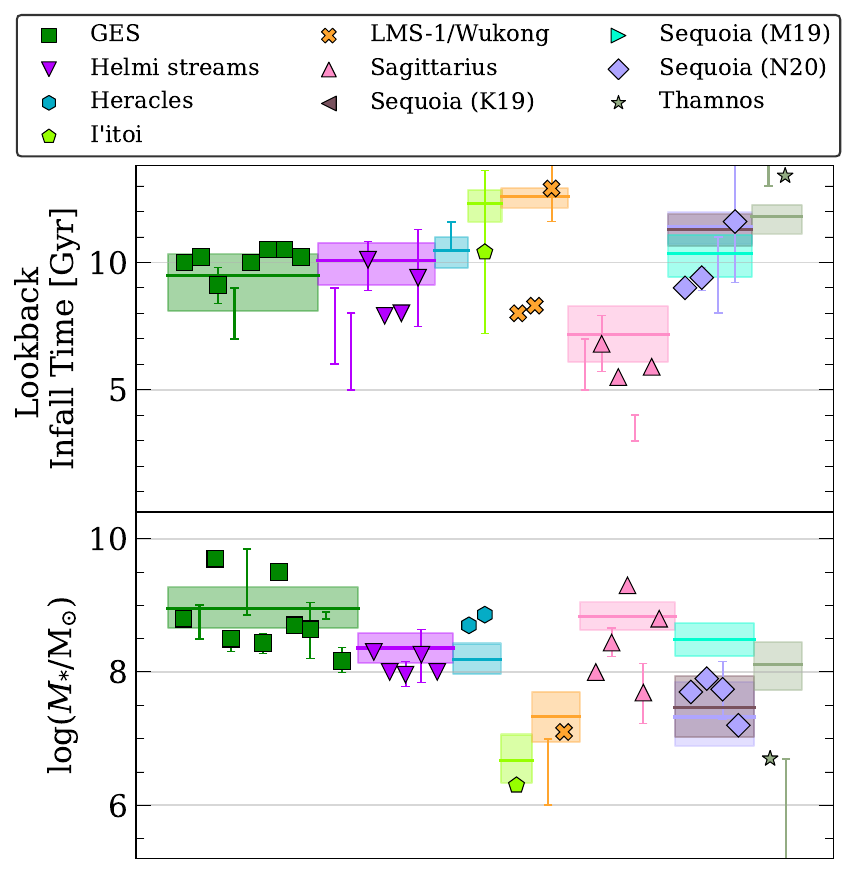}
    \caption{Comparison of the lookback infall time (top) and stellar mass (bottom) of the MW progenitors as predicted in this work (bands) and through other studies in the literature (scatter points and bars). Scatter points represent single-value estimates, whereas bars indicate estimates quoting a confidence interval. The estimates from the literature for the Sequoia are all represented with the same marker and colour corresponding to the {\it N20} sample, but should be compared to the predictions from the {\it M19} and {\it K19} samples, too.
    }
    \label{fig:comp_literature}
\end{figure*}

A comparison between the estimated lookback infall time and stellar mass of the MW progenitors as predicted by the GalactiKit methodology to the corresponding estimates in the literature is shown in the upper and lower panels of Fig.~\ref{fig:comp_literature}, respectively. The estimates from the inferred posterior distributions are reported as coloured bands, whose upper and lower limits represent the 84th and 16th percentiles of the distributions, respectively, while the continuous lines crossing the bands indicate the median values of the merger property. The lookback infall time and stellar mass predictions reported by other studies are also shown for both single value (scatter points) and confidence range (error bars) estimates. The colour and symbol of the bands, scatter points and error bars for each progenitor are consistent with what are used in Figs.~\ref{fig:IoM} and \ref{fig:alphairon}. References to the studies used for the comparison as well as the specific values of the lookback infall time and stellar mass of the progenitors shown in the figure are reported in Table~\ref{tab:comb_lit}. Note that these works differ from each other not only in terms of the methodologies adopted, but also in the definition of the infall time and stellar mass and in the way the samples of accreted stars from each progenitors are selected. To the best of our knowledge, our study is the first to provide self-consistent estimates for both the lookback time and stellar mass at infall for such a large number of MW disrupted progenitors. 

\begingroup
\renewcommand{\arraystretch}{1.3}
\begin{table*}
    \begin{tabular}{l|c c|c c}
        \hline
        & \multicolumn{2}{c|}{Lookback Infall Time [Gyr]} & \multicolumn{2}{c}{$\log (M_{*}\,/\,\text{M}_{\sun})$} \\
        \hline\\
        GES & $9.5^{+0.8}_{-1.4}$ & $\sim10$ \,\,(a) & $9.0^{+0.3}_{-0.3}$ & $8.8$ \,\,(q) \\
        & & $10.2^{+0.2}_{-0.1}$ \,\,(b) & & $8.5-9$ \,\,(r) \\
        & & $9.1^{+0.7}_{-0.7}$ \,\,(c) & & $9.7$ \,\,(s) \\
        & & $7-9$ \,\,(d) & & $8.5^{+0.1}_{-0.2}$ \,\,(t) \\
        & & $\sim10$ \,\,(e) & & $8.85-9.85$ \,\,(u) \\
        & & $10.5$ \,\,(f) & & $8.43^{+0.15}_{-0.16}$ \,\,(c) \\
        & & $10.5$ \,\,(g) & & $9.5$ \,\,(d) \\
        & & $10.2\pm0.2$ \,\,(h) & & $8.7$ \,\,(f) \\
        & & & & $8.64^{+0.40}_{-0.44}$ \,\,(v) \\
        & & & & $8.8-8.9$ \,\,(w) \\
        & & & & $8.16^{+0.21}_{-0.17}$ \,\,(x) \\
        & & & & \\
        
        Helmi streams & $10.1^{+0.7}_{-0.9}$ & $6-9$ \,\,(i) & $8.4^{+0.2}_{-0.2}$ & $8.3$ \,\,(y) \\ 
        & & $5-8$ \,\,(j) & & $\sim8$ \,\,(j) \\
        & & $10.1^{+0.7}_{-1.2}$ \,\,(c) & & $7.96^{+0.19}_{-0.18}$ \,\,(c) \\
        & & $7.9$ \,\,(f) & & $8.26^{+0.38}_{-0.42}$ \,\,(u) \\
        & & $\sim8$ \,\,(k) & & $8$ \,\,(f) \\
        & & $9.4\pm1.9$ \,\,(h) & & \\
        & & & & \\
        
        Heracles & $10.5^{+0.5}_{-0.7}$ & $10.5-11.6$ \,\,(l) & $8.2^{+0.2}_{-0.2}$ & $\sim8.7$ \,\,(l) \\
        & & & & $8.86^{+0.01}_{-0.01}$ \,\,(z) \\
        & & & & \\
        
        I'itoi & $12.3^{+0.6}_{-0.7}$ & $10.4\pm3.2$ \,\,(h) & $6.7^{+0.4}_{-0.4}$ & $6.3$ \,\,(f) \\
        & & & & \\
        
        LMS-1/Wukong & $12.6^{+0.4}_{-0.5}$ & $\sim8$ \,\,(m) & $7.3^{+0.4}_{-0.3}$ & $6-7$ \,\,(m) \\
        & & $8.3$ \,\,(f) & & $7.1$ \,\,(f) \\
        & & $12.9\pm1.3$ \,\,(h) & & \\
        & & & & \\
        
        Sagittarius & $7.2^{+1.1}_{-1.2}$ & $5-7$ \,\,(n) & $8.8^{+0.2}_{-0.2}$ & $\sim8$ \,\,(aa) \\
        & & $6.8^{+1.1}_{-1.1}$ \,\,(c) & & $8.8$ \,\,(f)\\
        & & $5.5$ \,\,(k) & & $8.44^{+0.22}_{-0.21}$ \,\,(c) \\
        & & $3-4$ \,\,(d) & & $9.3$ \,\,(d) \\
        & & $5.9$ \,\,(f) & & $7.69^{+0.43}_{-0.47}$ \,\,(u) \\
        & & & & \\
        
        Sequoia & $11.3^{+0.6}_{-0.6}$ (\textit{K19}) & $\sim9$ \,\,(o) & $7.4^{+0.5}_{-0.4}$ (\textit{K19}) & $\sim7.7$ \,\,(o) \\
        & $10.3^{+0.7}_{-0.9}$ (\textit{M19}) & $9.4^{+0.4}_{-0.5}$ \,\,(c) & $8.5^{+0.2}_{-0.3}$ (\textit{M19}) & $7.9^{+0.1}_{-0.1}$ \,\,(c) \\
        & $11.4^{+0.6}_{-0.7}$ (\textit{N20}) & $11-8$ \,\,(p) & $7.3^{+0.5}_{-0.4}$ (\textit{N20}) & $7.74^{+0.42}_{-0.46}$ \,\,(u) \\
        & & $11.6\pm2.4$ \,\,(h) & & $7.2$ \,\,(f) \\
        & & & & \\
        
        Thamnos & $11.8^{+0.5}_{-0.6}$ & $>13$ \,\,(p) & $8.1^{+0.4}_{-0.4}$ & $<6.7$ \,\,(cc) \\
        & & $13.4\pm0.3$ \,\,(h) & & $6.7$ \,\,(f) \\
        \hline
    \end{tabular}
    \caption{Lookback infall time and stellar mass predictions of the MW progenitors. From left to right, the columns indicate: the name of the progenitor accreted substructure (i), the lookback infall time and stellar mass predictions provided by GalactiKit (ii, iv) and studies from the literature (iii, v). 
    References: 
    a) \protect\cite{gallart_2019}, b) \protect\cite{bonaca_2020}, c) \protect\cite{Kruijssen_2020}, d) \protect\cite{Hasselquist_2021}, e) \protect\cite{montalban_2021}, f) \protect\cite{naidu_2022}, g) \protect\cite{gonzalez-koda_2025}, h) \protect\cite{Woody_2025}, i) \protect\cite{kepley_2007}, j/K19) \protect\cite{koppelman_2019}, k) \protect\cite{ruiz-lara_2022}, l) \protect\cite{horta_2021}, m) \protect\cite{Malhan_2021}, n) \protect\cite{deBoer_2015}, o/M19) \protect\cite{myeong_2019}, p) \protect\cite{dodd_2024}, q) \protect\cite{helmi_2018}, r) \protect\cite{Mackereth_2019}, s) \protect\cite{Vincenzo_2019}, t) \protect\cite{Mackereth_2020}, u) \protect\cite{feuillet_2020}, v) \protect\cite{Callingham_2022}, w) \protect\cite{han_2022}, x) \protect\cite{Lane_2023}, y) \protect\cite{helmi_1999}, z) \protect\cite{horta_2025}, aa) \protect\cite{Niederste-Ostholt_2012}, bb) \protect\cite{McConnachie_2012}, cc) \protect\cite{koppelman_2019_b}.}
    \label{tab:comb_lit}
\end{table*}
\endgroup

In general, the GalactiKit predictions for the infall times are in agreement with the estimates extrapolated through other methodologies for the most studied MW progenitors (GES, Helmi and Sagittarius galaxies). A tension in the literature is present for the infall time of the Helmi streams, where N-body simulations \citep{kepley_2007, koppelman_2019} and star formation history from CMD fitting \citep{ruiz-lara_2022} set the merger around 8~Gyr ago, while estimates from the associated globular cluster population \citep{Kruijssen_2020} and the ages distribution of field stars as determined through isochrone fitting \citep{Woody_2025} favour an earlier merger, around 10~Gyr ago. The GalactiKit prediction agrees with the latter scenario. 

Similarly, the GalactiKit predictions for the GES, Helmi and Sagittarius galaxies stellar masses are in agreement with results from the majority of previous studies. There is a significant variation among the reported stellar mass estimates of Sagittarius, which span the range of $10^{7.7}-10^{9.3}~\rm{M}_{\sun}$. GalactiKit predicts a stellar mass in the range $10^{8.6-9.1}~\rm{M}_{\sun}$ for the Sagittarius galaxy at accretion, which is similar to the estimate from the mass--metallicity relation of the globular cluster population associated with this system \citep{Kruijssen_2020}.

The nature of the Sequoia substructure \citep{myeong_2019, koppelman_2019_b} is a matter of debate in the literature, with some studies assigning a standalone origin as the remnant of a disrupted satellite galaxy \citep{myeong_2019, matsuno_2019, Gaucimara-bethencourt2023, ceccarelli_2024}, while others argue for a connection with the GES merger as a group of stars in high-energy, retrograde orbits \citep{koppelman_2019_b,koppelman_2020, amarante_2022} or a satellite of this system \citep{naidu_2021}. In our study, the predictions of the infall time and stellar mass of the Sequoia dwarf galaxy are calculated considering the debris samples obtained from three different selection criteria. If the selection criteria were equivalent, i.e., different methods to sample debris from the same substructure, the merger properties estimated by GalactiKit conditioned on each debris sample would be consistent with each other. This follows from the way GalactiKit was implemented as the training set contains multiple independent, identically distributed samples drawn from the debris distribution of each merger event in the Auriga simulations. However, we note a similarity between the predictions from the K19 and N20 debris samples which predict a $\sim10^{7}~\rm{M}_{\sun}$ merger occurred $\sim12~\rm{Gyr}$ ago. Both of these estimates are in disagreement with the one obtained from the M19 sample, which points toward a larger ($\sim10^{8}~\rm{M}_{\sun}$), more recent merger ($\sim11~\rm{Gyr}$ ago). Therefore, the time and stellar mass estimates from the \textit{K19} and \textit{N20} samples seem to point towards Sequoia being a separate MW progenitor than GES, while the estimates from the \textit{M19} selection could be consistent with the scenario of a substructure forming as a result of the GES merger. Interestingly, \citet{Horta_2023} also find a dichotomy in the origin of Sequoia as the chemical abundance patterns of the stars in the \textit{M19} sample are similar to those of GES, while the \textit{K19} and GES debris appear chemically inconsistent. This can be seen in Fig.~\ref{fig:alphairon}, which shows that the {\it M19} sample is characterised by a higher $\rm{[Fe/H]}$ and lower $\rm{[Mg/Fe]}$ population than the {\it K19} and {\it N20} samples, which could explain the higher stellar mass and later accretion time predictions for the Sequoia substructure in this case.

Stronger disagreements are present between the GalactiKit predictions and those from previous studies for the other substructures (Heracles, I'itoi, LMS-1/Wukong, Thamnos), which might be related to an incorrect physical description of low-mass systems ($M_{*}<10^{7}~\rm{M}_{\sun}$) in the Auriga simulations due to particle resolution limitations. While this might explain the discrepancies between the GalactiKit predictions and the values from literature for the progenitors of the I'itoi, LMS-1/Wukong and Thamnos substructures, the same is not true for the Heracles galaxy, whose stellar mass is more precisely estimated through density fitting as $10^{8.8}~\rm{M}_{\sun}$ \citep{horta_2025}. Another possible explanation is that the debris samples used to inform the GalactiKit model are not representative of the actual accreted dwarf galaxies. For instance, \citet{ceccarelli2025} conducted a high-resolution chemical abundance analysis on some of the stars associated with the Thamnos substructure finding a significant contamination from stars belonging to the MW in-situ population and, to a lesser degree, the GES galaxy. 

\subsection{The mass-metallicity relation of the MW progenitors}

Alongside the direct comparison with the infall time and stellar mass values from the literature, an implicit test of the validity of the GalactiKit predictions can be performed by investigating whether the predicted merger properties follow known physical relations. For this purpose, we compare the inferred progenitor stellar mass with the average measured metallicity of the stellar debris. Empirically, a positive relation is observed between these two quantities for galaxies in the mass range $10^{3.5-12}~\rm{M}_{\sun}$ \citep{kirby_2013}. This stellar mass -- metallicity relation (MZR) can be explained in terms of the amount of gas retained by galaxies through their gravitational pull. 

\begin{figure}
    \centering
    \includegraphics[width=\linewidth]{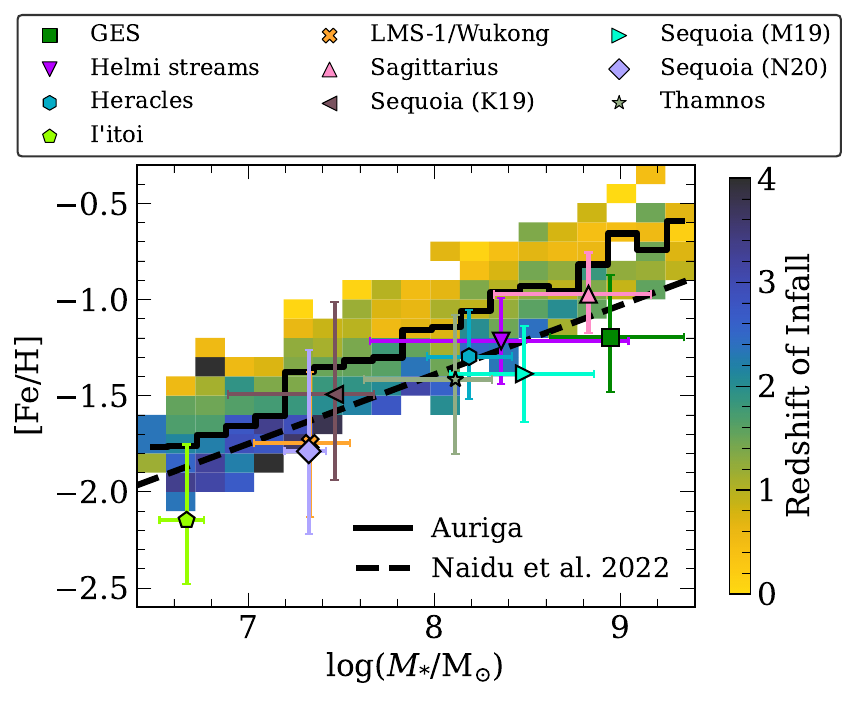}
    \caption{The mass--metallicity relation for the disrupted progenitors of the MW. Each progenitor is represented as a scatter point consistently with the scheme used in Fig.~\ref{fig:comp_literature}. The point is located at the median of the observed $\rm [Fe/H]$ debris distribution and the median of the inferred posterior distribution of the progenitor stellar mass. The extent of the errorbars shows the 16th-84th percentile range in each dimension. In the background, the distribution of the Auriga mergers is shown in mass--metallicity bins colour-coded by their average redshift of infall.  The continuous black line represents the median metallicity in each mass bin, while the dashed line shows the MZR inferred for the MW progenitors by \citet{Naidu_2020}. }
    \label{fig:MZR}
\end{figure}

Fig.~\ref{fig:MZR} shows the inferred MZR for the disrupted progenitors of the MW, represented as scatter points using the same scheme as in Fig.~\ref{fig:comp_literature}. The MZR in the Auriga simulations is also shown by binning the merger galaxies in both mass and metallicity. Random samples from the calibration model were added to the simulation data to allow a comparison with the observed metallicities. The colour of each bin indicates the median redshift at which the galaxies crossed the virial radius of the host. The continuous black line represents the median metallicity values in the mass bins. Another prediction for the MZR of the disrupted MW progenitors, inferred by \citet{naidu_2022} using comparisons with N-body simulations, is also shown as a dashed black line. In general, the GalactiKit stellar mass predictions are consistent with this previously predicted MZR. However, because progenitors were accreted at different redshifts, and because the MZR is expected to evolve with time, we generally expect the progenitors to follow different MZRs.  The time evolution of MZR is seen in both simulations \citep{ma_2016,torrey_2019, garcia_2024, grimozzi2024} and observations \citep{sanders_2021,roberts-borsani_2024}, a behaviour which is mainly due to a lower metal abundance of galaxies at higher redshifts. Thus, while the slope of the MZR tends to stay constant, the normalization factor of the relation scales over time. This can also be noticed by looking at the coloured bins in Fig.~\ref{fig:MZR}, which show that in the Auriga simulations galaxies with higher metallicity in a given mass bin have been accreted at earlier times. Therefore, galaxies from late merger events, such as Sagittarius, are expected to follow a MZR with higher normalization than galaxies accreted at earlier epochs (such as I'itoi and LMS-1/Wukong).  

\begin{figure}
    \centering
    \includegraphics[width=\linewidth]{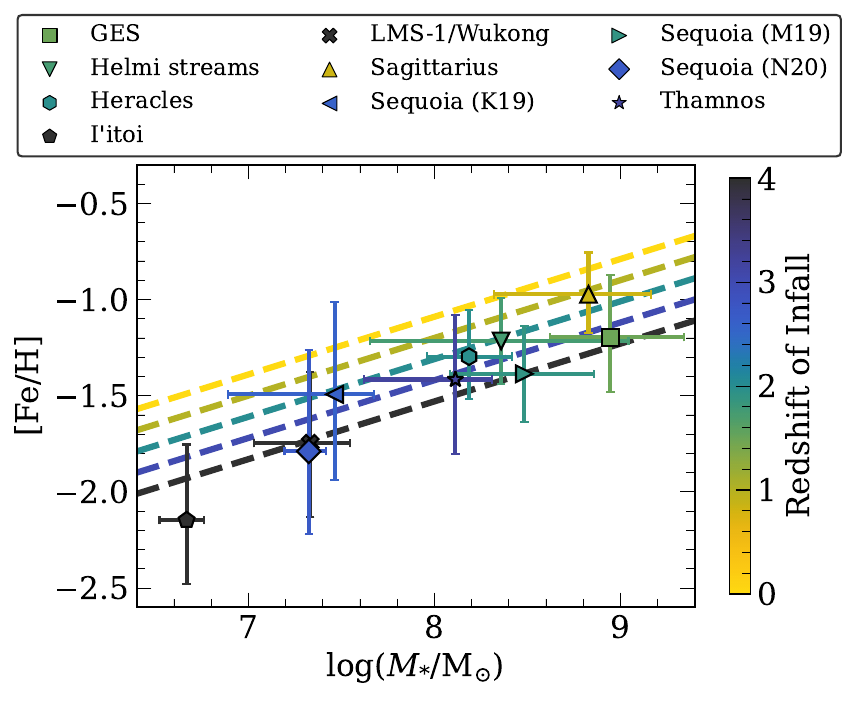}
    \caption{Mass--metallicity relation for the disrupted satellites of the MW as in Fig.~\ref{fig:MZR}. The dashed lines represent the empirical MZR calculated with equation~(\ref{eq:MZR}) for $z=0$ (top line), $z=1$, $z=2$, $z=3$ and $z=4$ (top to bottom), respectively.}
    \label{fig:MZR_obs}
\end{figure}

To investigate whether MW progenitors accreted at different times are consistent with different observed MZRs, we compute the redshift evolution of this relation by considering the MZR at $z=0$ \citep{kirby_2013} and modeling the change in the normalization factor as done in \citet{sanders_2021}, i.e., using a gradient of $d[{\rm{Fe}/{H}}]/dz\approx-0.11$. This evolution of the metallicity content comes from a study of the gas-phase metallicity of galaxies in redshift $z=0-3.3$ from the MOSDEF survey. Although some of the MW accreted progenitors are predicted to be accreted at $z>3.3$, recent JWST observations \citep{Nakajima2023,sarkar_2025} show that the MZR does not evolve significantly past this redshift. Therefore, we assume a constant ratio between the average stellar and gas-phase metallicity for galaxies of similar mass across redshifts and calculate the observed MZR as:

\begin{equation}
    \begin{aligned}
       \,[\rm{Fe/H}]_{\it z=z'} &  = -0.11\times z' + [\rm{Fe/H}]_{{\it z}=0} \\
        & =-0.11\times z' + (-1.69) + 0.30 \times \left( \log(M_{*}/\rm{M_{\odot}})-6 \right)
    \end{aligned}
    \label{eq:MZR}
\end{equation}
Fig.~\ref{fig:MZR_obs} shows the predicted MZR for MW progenitors against the same empirical relation calculated using equation~(\ref{eq:MZR}) at $z=0$ (top line), $z=1$, $z=2$, $z=3$ and $z=4$ (bottom line). The empirical lines are coloured by the redshift value at which they were estimated and, similarly, the predicted scatter points are coloured by the redshift of infall predicted for each merger. It appears that most of the MW accreted progenitors follow the MZR evaluated at the closest redshift to their time of infall. A discrepancy can be noticed between the different sample selections for the Sequoia galaxy. Although the \textit{K19} (left triangle) and \textit{N20} (diamond) stellar mass predictions are similar to each other, only the former lies on a MZR consistent with its predicted infall time. Similarly, the \textit{M19} (right triangle) stellar mass prediction, which is considerably larger than the one from the \textit{K19} and \textit{N20} samples, lies on a MZR inconsistent with its redshift of infall. Therefore, for the rest of the analysis, we consider the \textit{K19} sample as our preferred representation for the debris of the Sequoia galaxy because it also has a lower risk of contamination from GES stars compared to the more metal rich \textit{M19} sample.  

%

\subsection{The growth of the accreted stellar mass and halo mass of the MW}

The predicted lookback infall time and stellar mass of the MW progenitors allow us to trace back the evolution of the accreted mass content of the Galaxy across cosmic time. Because GalactiKit returns the probability distributions of these quantities, the cumulative distribution of the MW accreted stellar mass as a function of redshift, $\rm{CDF}_{MW}({\it{M}}_{\rm{*,ACC}},\it{z}\rm)$, can be estimated numerically from the ensemble of the single $\rm{CDF_{MW,\it{i}}}({\it{M}}_{\rm{*,ACC}},\it{z}\rm)$. A single ($i$-th) prediction of the MW assembly history can be computed by drawing a sample from the posterior distribution of each merger event. The cumulative MW accreted stellar mass distribution, $\rm{CDF_{MW,\it{i}}}({\it{M}}_{\rm{*,ACC}},\it{z}\rm)$,  can then be calculated by adding the stellar mass estimates of each event ordered by their redshift of infall, which is obtained from the lookback infall time assuming the same cosmology as in the Auriga simulations. We repeat this procedure for $i=1,000$ times obtaining a distribution of $\rm{CDF_{MW,\it{i}}}({\it{M}}_{\rm{*,ACC}},\it{z}\rm)$.  For this analysis, we exclude Thamnos, whose mass estimate is significantly over-predicted by the model compared to the estimates from the literature. 

\begin{figure}
    \centering
    \includegraphics[width=
    \columnwidth]{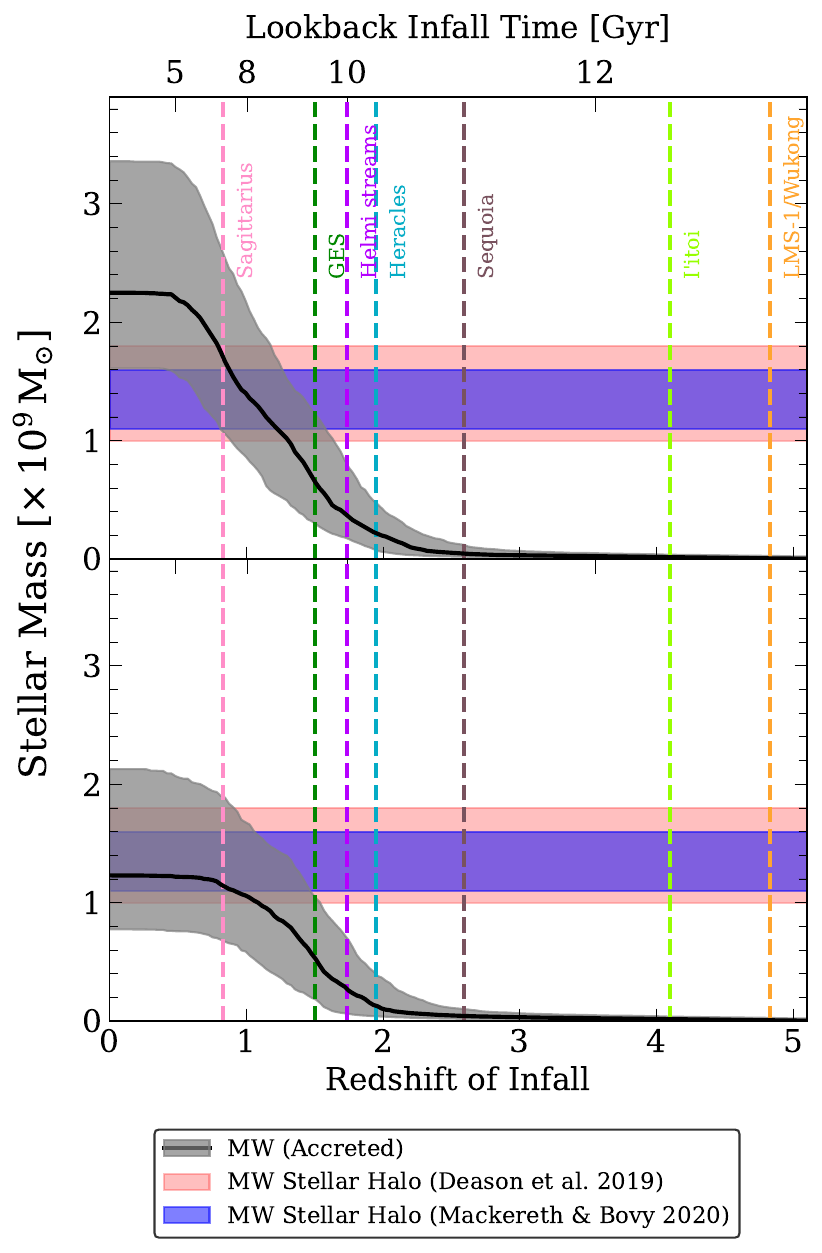}
    \caption{Cumulative accreted stellar mass distribution for the MW through cosmic time with (top) and without (bottom) the Heracles and Sagittarius mergers. This calculation only accounts for 1 Sequoia sample (\textit{K19}). The plot also reports measurements of the stellar mass of the MW halo, excluding the Magellanic Clouds, obtained through isochrone fitting \citep{Deason2019} and density modeling \citep{Mackereth_2020} of RGB stars.}
    \label{fig:MW_accreted_mass}
\end{figure}

Fig.~\ref{fig:MW_accreted_mass} shows the median (black line) and 16th-to-84th percentile range (grey area) of the predicted cumulative accreted mass distribution of the MW, $\rm{CDF}_{MW}({\it{M}}_{\rm{*,ACC}},\it{z}\rm)$. For reference, the median redshifts of infall of the considered merger events are shown as vertical dashed lines. The Galaxy is predicted to have obtained the majority of its accreted stellar mass between $z=1$ and $z=2$, following the mergers with the GES, Helmi streams and Heracles dwarf galaxies. A significant mass contribution is provided also by the latest merger with the Sagittarius dwarf galaxy, which in this calculation is considered in its totality, i.e., debris and surviving satellite. 

More quantitatively, the fraction of accreted stellar mass contributed by each merger event is estimated by calculating the ratio of the stellar mass of a progenitor and the total MW accreted stellar mass for each $i$-th assembly history. The median and 16th-to-84th percentile range of the distribution of the $1,000$ mass fractions of each merger event is reported in Table~\ref{tab:accreted mass}. GES is predicted to be the most significant contributor of accreted stars in the MW ($\sim0.43$), which may have contributed to the disc formation phases \citep{fukanoshi2025}. Only the Helmi streams ($\sim0.10$) and Heracles ($\sim0.07$) among the fully phase-mixed substructures contribute more than 0.01. The total stellar mass of Sagittarius is also a significant fraction ($\sim0.31$) of the mass supposedly accreted by the MW. However, because this is an ongoing merger, we cannot predict how much of this mass is still bound to the surviving satellite as GalactiKit predicts the mass of the system at the time it first interacted with the MW by crossing its virial radius.

\begin{table}
    \centering
    \begin{tabular}{l c c c}
    \hline
        & 50th & 84th & 16th \\
    \hline
       GES & 0.425 & 0.612 & 0.252\\
       Helmi streams  & 0.099 & 0.180 & 0.054 \\
       Heracles & 0.071 & 0.124 & 0.037\\
       I'itoi & 0.002 & 0.006 & 0.001 \\
       LMS-1/Wukong & 0.009 & 0.024 & 0.004 \\
       Sagittarius & 0.309 & 0.474 & 0.188 \\
       Sequoia & 0.013 & 0.042 & 0.005\\
       \hline
    \end{tabular}
    \caption{Fraction of progenitor stellar mass to MW total accreted stellar mass. From left to right, the columns indicate the median (50th), 84th and 16th percentile of the mass fractions MW progenitors estimated from 1,000 sampled realisations of the Galaxy assembly history.}
    \label{tab:accreted mass}
\end{table}

However, under the assumption that the majority of the accreted stellar content is within the stellar halo, an investigation of the amount of the accreted stellar mass retained by the MW after its mergers can be conducted comparing the predicted total accreted stellar mass to measurements of the stellar halo mass. For this purpose, we report in Fig.~\ref{fig:MW_accreted_mass} two estimates of the $z=0$ total stellar halo mass obtained from number counting \citep[red band,][]{Deason2019} and density modeling \citep[blue band][]{Mackereth_2020} of red giant branch (RGB) stars in the MW stellar halo. Neither estimates include the contribution of the Magellanic Clouds. The present-day MW accreted stellar mass predicted by GalactiKit is $2.2^{+1.1}_{-0.6} \times 10^{9}~\rm{M_{\sun}}$. This estimate is consistent within the uncertainty range with the stellar halo mass measured by \citet{Deason2019} ($1.4\pm0.4 \times 10^{9}~\rm{M_{\sun}}$), while being larger than the estimate from \citet{Mackereth_2020} ($1.3^{+0.3}_{-0.2} \times 10^{9}~\rm{M_{\sun}}$). However, the latter study does not account for the contribution of the Sagittarius dwarf galaxy \citep{Mackereth_2020}. Ignoring this contribution, as well as the one from the Heracles debris which occupies the inner region of the Galaxy \citep{horta_2021}, we found the total MW accreted stellar mass to be $1.3^{+1.0}_{-0.5} \times 10^{9}~\rm{M_{\sun}}$ (bottom panel of Fig.~\ref{fig:MW_accreted_mass}), which is in agreement with both studies. Therefore, the total stellar mass accreted from the progenitors of the substructures identified in the stellar halo is consistent with measurements of the mass of this component.  This suggests that, in line with other findings \citep[e.g., ][]{Naidu_2020}, the stellar halo is dominated by accreted stars.

Alongside providing predictions for the stellar mass and time of infall, GalactiKit infers also the halo mass of the accreted progenitor, $\log (M/\rm{M_{\sun}})$, and halo mass merger ratio,  $\rm{MMR}=\log(\it{M}/\rm{M}_{\sun})-\log(\it{M}_{\rm{MW}}/\rm{M}_{\sun})$, of each progenitor. This information can be combined to infer the MW halo mass, $\log (M_{\rm{MW}})$, at the merger time, i.e., 

\begin{equation}
    \log (M_{\rm{MW}})_{z=z_{\rm infall}} = \log(M_{\rm prog})_{z=z_{\rm infall}} - \rm{MMR}_{z=z_{\rm infall}}. 
    \label{eq:mw_mass}
\end{equation}

\begin{figure*}
    \centering
    \includegraphics[width=0.7
    \linewidth]{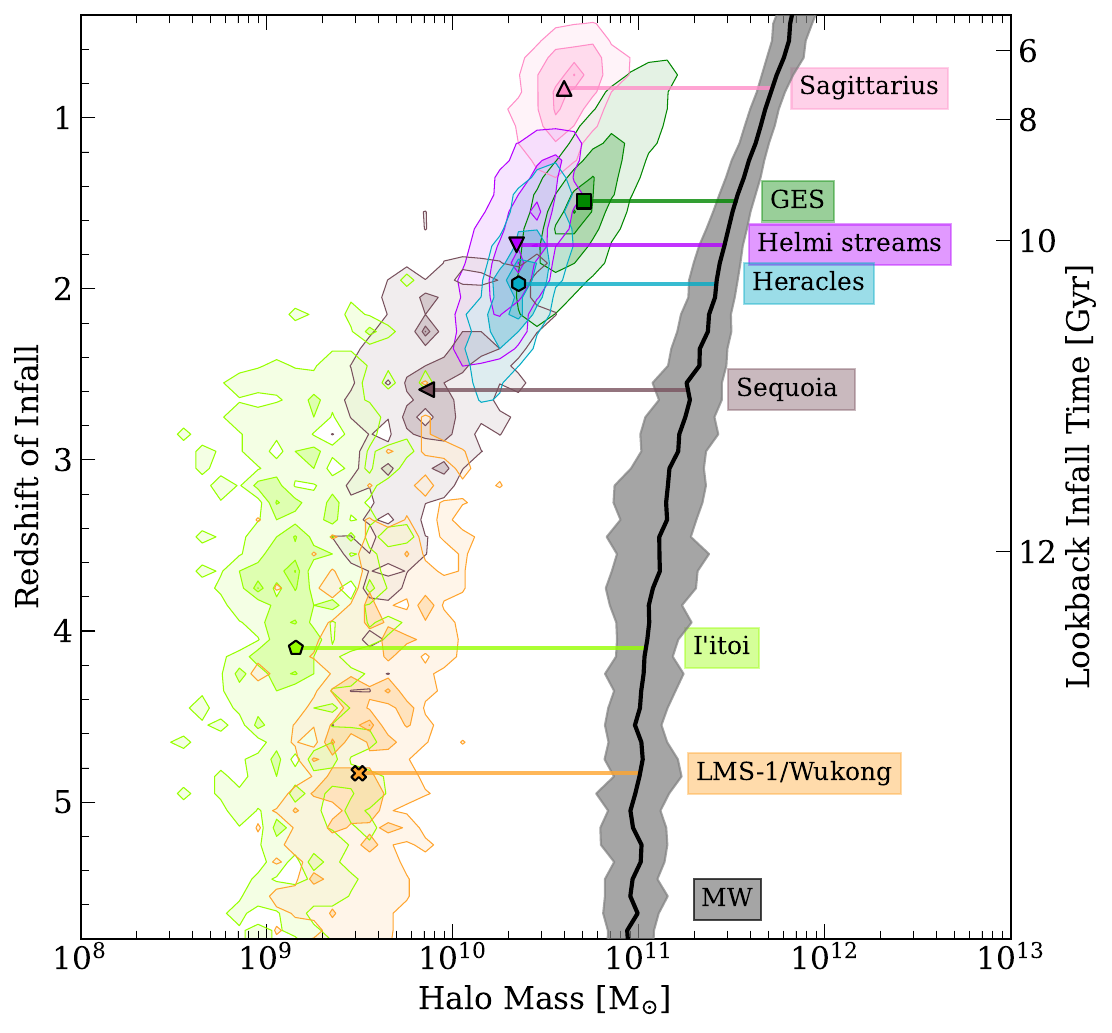}
    \caption{Dark matter halo mass growth of the MW as a function of redshift estimated from predictions of the progenitor halo mass  (coloured contours) and halo-mass ratio to host at infall. The contour levels represent $2\sigma$ constraints from the model. 
    }
    \label{fig:MW_halo_mass}
\end{figure*}

Therefore, the growth of the MW dark matter halo across cosmic time can be reconstructed from its assembly history. Given the underlying assumption that each merger is an independent event and that the progenitor properties are independent and identically distributed samples, a distribution of the MW halo mass across redshifts is calculated, using equation~(\ref{eq:mw_mass}), by combining the samples from all the progenitor posteriors. The median and 16th-to-84th percentile range of the MW halo mass estimates in different redshift bins is shown in Fig.~\ref{fig:MW_halo_mass} as a black line and a grey, shaded area. For context, we also show the density contours of the 2-D distribution of the halo mass and redshift of infall inferred for the progenitors considered in the calculation. Similarly to the previous analysis, we exclude samples from the Thamnos substructure.

At the earliest traced time ($z\approx5$), the Galaxy is predicted to live within a dark matter halo of mass $1.0^{+0.4}_{-0.3} \times 10^{11}~\rm{M_{\sun}}$.  After the first merger events (LMS-1/Wukong, I'itoi and Sequoia), the mass of the halo increases to $2.6^{+0.8}_{-0.8} \times 10^{11}~\rm{M_{\sun}}$ at $z\approx2$. This is unlikely due only to the accretion of those galaxies, whose dark matter halo masses are predicted to be $<10^{10}~\rm{M}_{\sun}$. A significant growth is observed after the accretion of the Heracles, Helmi streams and GES dwarf galaxies, as the Galaxy halo reaches a mass of $5.5^{+1.4}_{-1.1} \times 10^{11}~\rm{M_{\sun}}$ before starting its interaction with Sagittarius at $z\approx1$. It is interesting to see that GES contributed to a large mass increase around $z\approx1.5$, which may have led to the thick to thin disc formation transition due to the cold to hot gas accretion phase change \citep{Grand2018,fukanoshi2025}. Although the curve in Fig.~\ref{fig:MW_halo_mass} appears to suggest that the Galaxy grows after each merger, which is physically expected, this cannot be explicitly predicted by GalactiKit, which infers the properties of the MW-progenitor system at the time of infall. However, the post-merger total dark matter mass of the MW can be estimated by adding the contribution of the Sagittarius dark matter halo to the estimate of the Galaxy halo at the time of infall. Thus, the post-merger estimate of the MW dark matter halo is $5.9^{+1.4}_{-1.1} \times 10^{11}~\rm{M_{\sun}}$, which appears to be already in agreement with the lower end of the expected range from dynamical measurements of the total MW mass \citep[e.g.,][]{Wang2022}. However, this estimate does not account for the evolution of the Galaxy after $z\approx1$, which includes significant accretion events such as the merger with the Magellanic Clouds.


\section{Conclusions}
\label{sec:conclusions}

In this study, we presented simulation-based predictions of the lookback infall time, stellar mass, halo mass and halo mass merger ratio of the MW accreted progenitors. Applying the GalactiKit simulation-based inference framework to Auriga simulations, we developed a model of the posterior distribution of the properties of mergers at infall conditioned to the chemo-dynamical distribution of debris at $z=0$. However, we modified the original methodology of \cite{galactikit} to explicitly account for the discrepancy between stellar properties in simulations and observations (Section~\ref{sec:noise_model}), hence ensuring predictions are robust against this uncertainty. Moreover, an additional data aggregation step (Section~\ref{sec:fishnets}) was added to maximise the amount of information extracted from the stellar properties. We applied the new version of GalactiKit on a sample of unseen merger-debris examples from the Auriga simulations, finding that the posterior predictions are accurate and consistent with the expected distributions (Figs.~\ref{fig:validation_scatter} and \ref{fig:validation_tarp}). We then proceeded to estimate the properties of the MW mergers conditioned on the chemo-dynamics of their $z=0$ debris. Our main findings are:

\begin{enumerate}
    \item a general agreement exists between the GalactiKit predictions and values from the literature for the lookback infall time and stellar mass of the MW progenitors (Fig.~\ref{fig:comp_literature}). A notable exception is the Thamnos substructure, whose stellar mass is considerably larger than the value reported by \citet{koppelman_2019_b}. However, this can be caused by significant contamination by in-situ stars in the merger debris sample \citep{ceccarelli2025}.
    \item the infall time and stellar mass predictions obtained from samples of the Sequoia progenitor obtained from different selection criteria are inconsistent with each other. The \textit{M19} sample predicts a merger event similar in time and mass to the accretion of GES, while the \textit{K19} and \textit{N20} samples point towards an earlier and less massive merger. This might suggest that the three selection criteria used here may not fully represent this system, or that there is a significant contamination from GES stars in the \textit{M19} sample. 
    \item the predicted progenitor stellar masses are positively related to the debris median metallicities, with mergers following different relations depending on their redshift of infall (Fig.~\ref{fig:MZR_obs}). 
    \item the predicted total stellar mass accreted by the MW is $2.2^{+1.1}_{-0.6} \times 10^{9}~\rm{M_{\sun}}$ (Fig.~\ref{fig:MW_accreted_mass}), with GES ($\sim0.43$) and Sagittarius ($\sim0.31$) accounting for  the majority of it. Excluding the mass contributions from Heracles and Sagittarius, we predict a total accreted stellar mass of $1.3^{+1.0}_{-0.5} \times 10^{9}~\rm{M_{\sun}}$ from substructures in the stellar halo. This is in good agreement with the mass measurements of this component of the Galaxy.
    \item the inferred growth of the MW dark matter halo mass is not entirely explained by the current census of mergers, suggesting the potential presence of undetected accretion events, especially between $z\approx5$ and $z\approx2$. The prediction for the MW halo mass after the accretion of Sagittarius
    is $5.9^{+1.4}_{-1.1} \times 10^{11}~\rm{M_{\sun}}$.  
\end{enumerate}

We caution that our results rely strongly on the selection of the debris samples defining each accreted progenitor and that GalactiKit assumes that each sample conditioning the merger property posterior is drawn independently from the complete chemo-dynamical distribution of the debris. Hence, the impact of the contamination from the in-situ population or from stars accreted from other mergers is not accounted for. This is a scope for future investigations. Similarly, potential biases caused by the surveys selection functions affecting the observable debris chemo-dynamical distribution remain to be studied. In addition, the choice of particle resolution in simulations has an effect on the realism of small galaxies and, consequently, on the predictions of their properties at infall. A comparison between the infall time and stellar mass predictions for these systems obtained from posterior estimators trained at different resolution levels is needed to quantify the importance of resolution effects on SBI predictions of merger events.

Nevertheless, we find that the SBI methodology provides remarkable insights on the Galaxy formation history and that the accuracy of these results is on a par with those from the equivalent corpus of studies in the literature. Modern and upcoming surveys provide a unique opportunity to advance these machine learning techniques, unlocking the potential to reconstruct the formation of the Galaxy and the Local Group with unprecedented fidelity.


\section*{Acknowledgements}

AS acknowledges a Science Technologies Facilities Council (STFC) Ph.D. studentship at the LIV.INNO Centre for Doctoral Training ``Innovation in Data Intensive Science''. This study used the Prospero high performance computing facility at Liverpool John Moores University. AS thanks Danny Horta for insightful discussions on the MW formation history. AF acknowledges support from UKRI (ST/W006766/1). RJJG acknowledges support from an STFC Ernest Rutherford Fellowship (ST/W003643/1). DK acknowledges MWGaiaDN, a Horizon Europe Marie Sk\l{}odowska-Curie Actions Doctoral Network funded under grant agreement no. 101072454 and also funded by UK Research and Innovation (EP/X031756/1). TLM acknowledges support from the InfoSys Physics-AI Lab at the University of Cambridge.

\section*{Data Availability}
The Auriga simulations can be downloaded through the Globus file transfer service (\url{https://globus.org/}). Detailed instructions on how to access the data are provided in \url{https://wwwmpa.mpa-garching.mpg.de/auriga/data.html}. 

The codes and models developed in the analysis are available under reasonable request to the authors. 



\bibliographystyle{mnras}
\bibliography{example} 





\bsp	
\label{lastpage}
\end{document}